\documentclass[sn-mathphys]{sn-jnl}% Math and Physical Sciences Reference Style
\usepackage{subfig}
\jyear{2021}%

\theoremstyle{thmstyleone}%
\theoremstyle{thmstyletwo}%

\theoremstyle{thmstylethree}%

\begin{document}

\title[Comparative study of ML and deep learning methods on ASD classification]{Comparative study of machine learning and deep learning methods on ASD classification}

%%=============================================================%%
%% Prefix	-> \pfx{Dr}
%% GivenName	-> \fnm{Joergen W.}
%% Particle	-> \spfx{van der} -> surname prefix
%% FamilyName	-> \sur{Ploeg}
%% Suffix	-> \sfx{IV}
%% NatureName	-> \tanm{Poet Laureate} -> Title after name
%% Degrees	-> \dgr{MSc, PhD}
%% \author*[1,2]{\pfx{Dr} \fnm{Joergen W.} \spfx{van der} \sur{Ploeg} \sfx{IV} \tanm{Poet Laureate} 
%%                 \dgr{MSc, PhD}}\email{iauthor@gmail.com}
%%=============================================================%%

\author*{\fnm{Ramchandra} \sur{Rimal}}\email{ramchandra.rimal@mtsu.edu}

\author{\fnm{Mitchell} \sur{Brannon}}\email{mb2at@mtmail.mtsu.edu}
%\equalcont{These authors contributed equally to this work.}

\author{\fnm{Yingxin} \sur{Wang}}\email{yw3n@mtmail.mtsu.edu}

\author{\fnm{Xin} \sur{Yang}}\email{Xin.Yang@mtsu.edu}
%\equalcont{These authors contributed equally to this work.}

\affil*{\orgdiv{Department of Mathematical Sciences}, \orgname{Middle Tennessee State University}, \orgaddress{\street{612 Old Main Cir KOM}, \city{Murfreesboro}, \postcode{37132}, \state{TN}, \country{USA}}}

%\affil[2]{\orgdiv{Department}, \orgname{Organization}, \orgaddress{\street{Street}, \city{City}, \postcode{10587}, \state{State}, \country{Country}}}

%\affil[3]{\orgdiv{Department}, \orgname{Organization}, \orgaddress{\street{Street}, \city{City}, \postcode{610101}, \state{State}, \country{Country}}}

%%==================================%%
%% sample for unstructured abstract %%
%%==================================%%

\abstract{The autism dataset is studied to identify the differences between autistic and healthy groups. For this, the resting-state Functional Magnetic Resonance Imaging (rs-fMRI) data of the two groups are analyzed, and networks of connections between brain regions were created. Several classification frameworks are developed to distinguish the connectivity patterns between the groups. The best models for statistical inference and precision were compared, and the tradeoff between precision and model interpretability was analyzed. Finally, the classification accuracy measures were reported to justify the performance of our framework. Our best model can classify autistic and healthy patients on the multisite ABIDE I data with 71\% accuracy.}


%%\pacs[JEL Classification]{D8, H51}

%%\pacs[MSC Classification]{35A01, 65L10, 65L12, 65L20, 65L70}

\maketitle

\section{Introduction}
\label{intro}
Brain Imaging Analysis is one of the emerging fields in cognitive neuroscience. Imaging studies reveal insights about normal brain function and structure, neural processing and neuroanatomic manifestations of psychiatric and neurological disorders, and neural processing alterations associated with treatment response~\cite{Bowman2014BrainIA}. This study performs the imaging methods to identify Autism Spectrum Disorder (ASD). ASD is a developmental disability that may cause difficulties in communicating and interacting with others and challenge their learning ability. Children with ASD tend to exhibit repetitive behavior or fixate on a particular interest. ASD is one of the fastest-growing developmental disorders that affect communication and behavior in the United States~\cite{thabtah2018new}. The advancements in medicine and technology have allowed physicians to assess, diagnose, and treat the complexities of ASD utilizing non-invasive techniques. Specifically, advancements in Brain Imaging Analysis have given physicians and researchers a more robust understanding of brain development. Studies conducted have shown, using Brain Imaging Analysis, that those with ASD have altered brain connectivity due to the nature of accelerated growth in the brain during development~\cite{lord2018autism}. According to the Centers for Disease Control and Prevention (CDC) data, the occurrence of diagnosis has increased over the years as 1 in 54 children are diagnosed with ASD. Unfortunately, achieving an accurate diagnosis is not always easy because it is based on heterogeneous behavior rather than measuring biological markers. Furthermore, there is no cure for ASD, but the most widely accepted forms of effective treatments include behavioral interventions~\cite{lord2018autism}. Emphasizing early, accurate diagnosis can contribute to successful, albeit varying levels of success, patient outcomes due to earlier treatment interventions.

A popular tool used in neuroimaging is Functional Magnetic Resonance Imaging (fMRI) which looks at the magnetic properties given by fluctuating levels of oxyhemoglobin and deoxyhemoglobin, which are affected by fluctuating levels of neural activity. Thus, the fMRI detects this Blood Oxygen Level Dependent (BOLD) signal to create a 3D image of the brain. This has given physicians and researchers a deeper understanding of how brain activity works, and the implications on ASD are significant. fMRI is very popular because it is a non-invasive method to obtain neural activity information from the human brain. Currently, two modalities of the fMRI have been used, the rs-fMRI and task-based fMRI. In this article, we work with the rs-fMRI data. The rs-fMRI is a neuroimaging tool that measures spontaneous low-frequency fluctuations in the neural BOLD signal of a subject at rest to investigate the functional architecture of the brain~\cite{khosla2019machine}. Functional Connectivity (FC), a pairwise relationship between two brain regions, is considered an essential step in searching for neuro markers in the ASD subject. The functional organization of networks involved in social and emotional processing is different between autism and healthy individuals; thus, rs-fMRI can be used as a diagnostic tool for ASD~\cite{kennedy2008intrinsic}. Several supervised and unsupervised learning methods have been used in analyzing the rs-fMRI data based on FC. Unsupervised machine learning is suitable for analyzing high-dimensional data with complex structures \cite{khosla2019machine}. Meanwhile, several supervised learning methods have been demonstrated to assess the mapping between input features and corresponding target predictions. Since systematic alterations in resting-state patterns are reported to be associated with pathology or cognitive traits, analyzing rs-fMRI data using supervised methods could facilitate the accuracy of ASD diagnostics~\cite{wang2019functional}. Khan et al. (2020) \cite{khan2020three} implemented the sequential forward feature selection approach to obtain highly distinguishing features between autistic and healthy controls. The results in~\cite{assaf2010abnormal} show a decreased FC among the ASD subjects as compared to the controlled subjects. In the task unrelated neuronal activity between 23 ASD and 20 control subjects, the study in~\cite{jones2010sources} found the disruptions in FC. The work on \cite{dadi2019benchmarking} focused on the model interpretability. They implemented linear classification techniques to classify not only ASD but also Alzheimer's and Schizophrenia. \cite{Parikh2019} implemented several classification techniques to classify ASD and normal patients and concluded that the neural networks outperformed other methods.

 While the existing diagnostic methods for ASD have high accuracy for homogeneous and small data sets, the classification capacity is not sufficient for heterogeneous and multisite data \cite{dvornek2017identifying}. Due to the complex nature of ASD and the difficulties presented in diagnosis, we seek to develop a more effective classification framework to improve the accuracy of ASD classification for heterogeneous and multisite data. Furthermore, we work on the classification of ASD based on FC, emphasizing the comparative study for interpretability of the model and the accuracy of the results.

\section{Literature Review}
Current machine learning diagnostic frameworks face challenges due to the heterogeneous nature of ASD data. This data derives information from fMRI and diffusion MRI, which may not directly capture relational, morphological changes between brain regions. Therefore,~\cite{bilgen2020machine} performed crowdsourcing to create a pool of machine learning pipelines for the diagnosis of ASD using cortical morphological networks (T1-weighted MRI) in Kaggle competition. All methods laying the foundation of those pipelines were examined under three machine learning steps: preprocessing techniques, dimensionality reduction methods, and learning models. A total of 20 teams participated in the study, and their performances were ranked after evaluating the accuracy, sensitivity, and specificity metrics of their models. The first-ranked team achieved an accuracy of 0.7, the sensitivity of 0.725, and the specificity of 0.675, respectively, demonstrating its discriminative potential in diagnosing ASD. Wang et al.~\cite{wang2019functional} implemented a support vector machine recursive feature elimination (SVM-RFE) method, a FC-based algorithm, to distinguish ASD from the control group. This study involved 255 ASD patients and 276 TD controls from 10 settings. Based on the social motivation hypothesis, 35 regions of interest were selected to construct the FC matrix. First, SVM-RFE was conducted to screen bioinformatics in the complex high-dimensional FC dataset using a stratified, 4-fold cross-validation approach. Then the selected features were imported into the SVM with a Gaussian kernel for classification. The developed classification algorithm appeared to improve the classification accuracy for the overall test and the leave-one-site-out test. These results indicated the new classification algorithm's capacity to accurately measure the importance of features and select the subset of most discerning features, which was superior to the current similar studies. \\

In the paper, Dvornek et al.~\cite{dvornek2017identifying}, the authors aimed to distinguish the ASD individuals from resting-state functional MRI time-series using long short-term memory (LSTM) networks. Utilizing the standardization of the time scale and data augmentation, 11000 input sequences from 1100 subjects consisting of 539 individuals with ASD and 573 typical controls from 17 international sites were analyzed in this study. The LSTM weights were applied to the data to account for both the current state and the signal from previous states; after stratified 10-fold cross-validation, the LSTM with 32 hidden nodes was found to achieve a classification accuracy of 68.5\%. This model was demonstrated to provide high classification accuracy on large heterogeneous datasets, which opens the door for further studies to investigate the mechanism of ASD. Kong et al.~\cite{kong2019classification} looked at standard classification methods used for ASD and sought to construct a deep neural network (DNN) classifier to diagnose ASD more efficiently. The researchers looked at small fMRI data consisting of 182 subjects, 78 of whom had ASD. They constructed an individual network for each subject utilizing the Destrieux atlas for parcellating different brain regions. Then the features that are extracted from the regions of interest (ROIs) are ranked using an F-score, which measures the discrimination of two sets of real numbers via the DNN classifier. Their classification method achieved about 90.39\% accuracy, which was significantly higher when compared to other existing methods. The researchers repeated their experiment with another dataset, resulting in an 86.70\% accuracy. Therefore, the researchers concluded that their proposed method was effective in ASD classification. It should be noted; however, the researchers looked at homogenous data sets juxtaposed to other research that utilized heterogeneous data sets. Heinsfeld et al.~\cite{heinsfeld2018identification} wanted to accurately classify ASD patients, based on their brain activation patterns alone, by applying deep learning algorithms. The dataset used was rs-fMRI data from the Autism Imaging Data Exchange, which included 1,035 patients, 505 of whom were ASD patients. The study used two stacked denoising autoencoders to extract the information of the dataset. The results were compared to the Support Vector Machine (SVM) and the Random Forest (RF) models. The researchers of this study were able to achieve an accuracy of 70\%, which was the highest classification achieved in comparison to the SVM and RF. The study concluded that deep learning methods might be a reliable way to classify ASD, despite data collected from different equipment and demographics. However, the study falls short of biomarker standards despite taking steps in the right direction towards more reliable results. \\

Anirudh et al.~\cite{Anirudh8683547} proposed a new approach to generating sets of population graphs predictive modeling, which was a bootstrap version of graph convolutional neural network (G-CNNs). The authors used weakly trained G-CNNs and reduced the sensitivity of the model selection graph structure. The dataset used consists of 1112 patients of the Autism Brain Imaging Data Exchange (ABIDE) dataset. Results indicated that the accuracy of novel method prediction accuracy was 70.86\%, which is superior to the G-CNNs's 69.50\%. In the article, Thomas et al.~\cite{thomas2020classifying} trained a full three-dimensional convolutional neural network (3D-CNN) on a cohort of about 2000 cases of Autism Brain Imaging Data Exchange (ABIDE)-I and II] data set. In this method, the authors achieved an accuracy of 66\%, which was comparable to the single measure regional homogeneity. In addition, they further applied the SVM method on ABIDE, which achieved an accuracy of 60\%, suggesting that 3D-CNNs could not detect additional information from these temporal transformations that were more useful to recognize ASD from controls. The paper by Gazzar et al.~\cite{el2019simple} introduced the graph convolution networks (GCN), which was a semi-supervised deep learning approach for using rs-fMRI to predict the diagnosis of ASD. The authors implemented a tenfold cross validation and across-sites cross validation scheme on the 1-D convolutional network. On the combined ABIDE I and II datasets, an accuracy of 0.68 and 0.65 were achieved, respectively.\\

The study by Sherkatghanad et al.~\cite{sherkatghanad2020automated} seeks to build an automated framework that can accurately identify and classify ASD patients and control subjects. The architecture that was proposed was a convolutional neural network (CNN); however, the researchers also considered other learning methods such as SVM, K-nearest neighbors (KNN), and RF. Their proposed CNN framework achieved a 70.22\% accuracy score on the ABIDE I rs-fMRI dataset.
In this article, Li et al.~\cite{li20182} examined biomarkers to identify characteristics of ASD and proposed a novel whole-brain fMRI-analysis scheme to discriminate ASD. C3D convolution architecture was developed to capture 3D spatial features \cite{tran2015learning}. Since fMRI was comprised of temporal and spatial information, the authors explored the temporal statistics with the sliding window method. They also studied the spatial features using a 2-channel convolutional 3D deep neural network (2CC3D). The proposed method generated an F-score of $0.89 \pm 0.05$, a mean of F-score improved over 8.5\% compared to traditional machine learning models.
Subbaraju et al.~\cite{subbaraju2017identifying} proposed a spatial feature-based detection method for feature identification of ASD by using rs-fMRI for accurate diagnosis of ASD. To examine the performance for ASD diagnosis, the authors conducted a detailed study using the large-scale ABIDE dataset, stratified by different gender and age groups. As a result, the SFM method could detect ASD in 78.6\% for adolescent males, 85.4\% for adult males, 86.7\% for adolescent females, and 95\% for adult females, which was superior in comparison to other methods in the existing literature. \\ 

Wang et al.~\cite{wang2020aimafe} proposed an ASD identification approach based on multi-atlas deep feature representation and ensemble learning. First, the authors calculated multiple FCs between each pair of regions based on different brain atlases from fMRI data of each subject and extracted these FCs as the original features. Then, to better classify ASD, they applied the stacked denoising autoencoder to perform the multi-atlas deep feature representation. Finally, they proposed a multilayer perceptron and an ensemble learning method to perform the final ASD identification task. The model was validated using a total of 949 patients (including 419 ASDs and 530 typical controls) from the ABIDE-I dataset. The authors achieved an accuracy of 74.52\%, a sensitivity of 80.69\%, specificity of 66.71\%, and AUC of 0.8026, respectively, which demonstrated its discriminative potential in diagnosing ASD. Almuqhim et al.~\cite{almuqhim2021asd} devised a deep-learning model named ASD-SAENet for ASD classification using fMRI data. The sparse autoencoder (SAE) is an unsupervised machine learning method that results in the optimized extraction of features for classification. Subsequently, feeding extracted features into a deep neural network leads to a superior classification of fMRI brain scans that are more prone to ASD. ASD-SAENet model is trained to optimize the classifier while improving extracted features based on reconstructed data error and the classifier error. They evaluated the ASD-SAENet model to learn features from the large-scale ABIDE-I dataset. The results indicated that the ASD-SAENet model developed an accuracy of 70.8\%, the sensitivity of 62.2 \%, and specificity of 79.1\% for the ABIDE-I dataset, which was superior to other methods in the existing literature.
However, phenotypic features carry predictive information and are always available; they were less likely to be included in the model. To overcome this problem, six distinct classification approaches, including Phenotype-TS, RawPhenotype-LSTM, EncPhenotype-LSTM, RawPhenotype-rsfScore, EncPhenotype-rsfScore, and Phenotype-Target, were implemented by Dvornek et al.~\cite{dvornek2018combining} to incorporate phenotypic data with rsfMRI into a single deep learning framework for classifying ASD. They tested the proposed architectures using a cross-validation framework on the cohort of ABIDE-I (including 527 ASDs and 571 typical controls). Among all developed models, the RawPhenotype-rsfScore, D = 2 method outperformed prior work by achieving a mean accuracy of 67.9\% and a mean subject accuracy of 70.1\%.
%%%%%%%%%%%%%%%%Section 2 Literature review %%%%%%%%%%%%%%%%%%%%%%%
\section{Methods}
\label{Methods}
The rest of the paper is organized as follows. We start with providing a computational framework for this work in section~\ref{Computational_Framework}. Then, a brief review of the LSTM and GRU neural networks with the pseudo-code for their implementation  is provided in section~\ref{Methods_Review}. Then the data collection and preprocessing steps are discussed in section~\ref{Methods_Data-preprocess}. Next, the methods for generating functional connectivity between the brain regions of interest are presented in section~\ref{Methods_fun-connectvitiy}. Next, the implementation of principal component analysis (PCA) for dimension reduction is described in section~\ref{Methods_dim_red}. The discussion on the experimental methods and the hyperparameter tuning procedure is discussed in section~\ref{experimental_methods} and section~\ref{Hyperparameter Tuning}, respectively. Next, the detailed results on the model performance are reported in section~\ref{Methods_model_performance} respectively. In addition, the conclusion and discussion of results are presented in section~\ref{discussion}. After the acknowledgment, the references used in this article are provided in the references section. Finally, the hyperparameter tuning results are provided in  the appendix section, where section~\ref{ Hyperparameter tuning for single layer LSTM on various RSFCs} and section~\ref{Hyperparameter tuning for single layer GRU with various RSFCs} contains all the results for the hyperparameter tuning for LSTM and GRU methods on all RSFCs, respectively.% ~\ref{reference_section}.%%%%%%%%%%%%%%%%%%%%%%
%The section~\ref{Methods_Review_LSTM} and  section~\ref{Methods_Review_GRU} provides the brief overview of LSTM neural networks and the GRU neural networks respectively.
%
\subsection{Computational Framework}
\label{Computational_Framework}
\begin{figure}[h]
  \includegraphics[scale=.5]{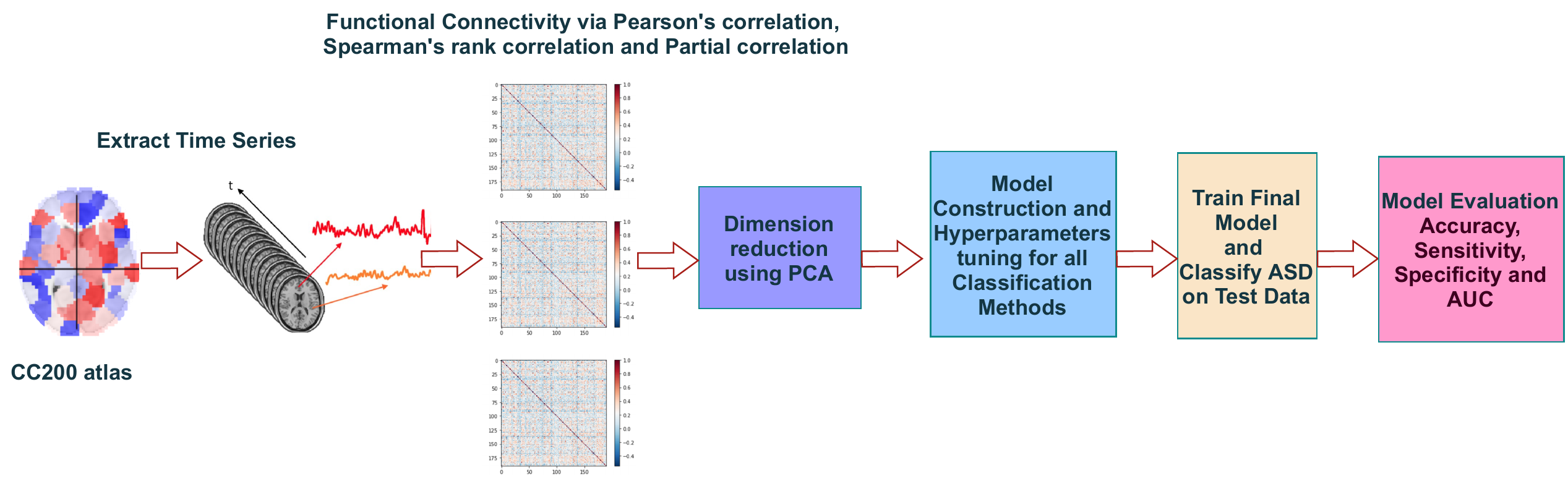}
 \caption{{\bf Schematic diagram of proposed model framework}}
 \label{Figure:Proposed_Framework}
\end{figure}

The proposed model framework is presented via the schematic diagram in Figure~\ref{Figure:Proposed_Framework}. As outlined in the diagram, the proposed research starts with extracting the time series from the CC200 atlas proposed in~\cite{craddock2012whole}. Then the RSFCs were created using Pearson's correlation, Spearman's rank correlation, and the Partial correlation. The upper triangular elements of those connectivity matrices were extracted as the features. The dimension reduction technique, known as  PCA, is implemented, and 871 subjects with 600 features were selected for the modeling. The classification models are constructed for all classification methods experimented, and the best value of hyperparameters is obtained. Next, using the best value of hyperparameters chosen, the final model is fit on the training data, and the model is applied to classify the ASD and healthy patients on the test data set. Finally, the performance of our model is evaluated using various measures such as accuracy, sensitivity, specificity, and AUC. The details of the procedures are provided in the corresponding sections and subsections.
%

%%%%%%%%%%%%%%%%%%%%%%%%%%%%%%
\subsection{Review of Classification Techniques}
\label{Methods_Review}
This article experimented with several supervised learning methods such as Logistic Regression (LR), Linear Support Vector Classifier (LSVC), Kernel Support Vector Classifier (KSVC), Random Forest Classifier(RFC), AdaBoosting classifier (ABC), Gated Recurrent Unit (GRU), and LSTM neural networks. However, since LR, LSVC, KSVC, RFC, and ABC are standard classification techniques widely available in any popular statistics textbooks~\cite{james2013introduction, friedman2001elements, kuhn2013applied}, an overview of these methods is not provided. So we provide the review of the neural networks based classification methods LSTM and GRU below.
%%%%%%%%%%%%%%%%%%%%%%%%
\subsection*{A brief overview of LSTM}
\label{Methods_Review_LSTM}
\begin{figure}[h]
   \includegraphics[scale=.32]{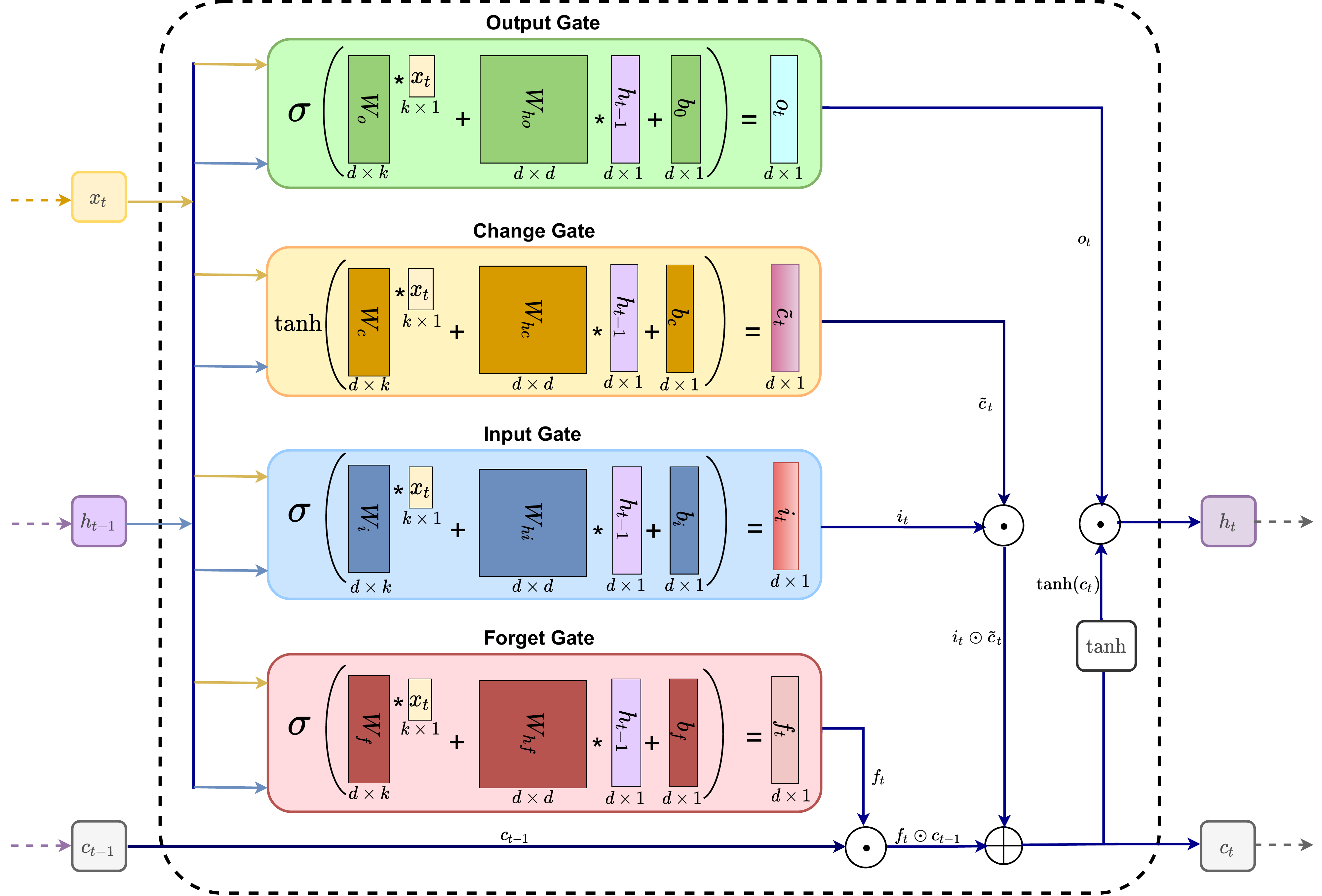}
  \caption{{\bf Long short-term memory(LSTM) architecture~\cite{bhandari2022predicting}.}}
  \label{Figure:LSTM_architecture}
\end{figure}

 LSTM and GRU  are popular and more recent variants of recurrent neural networks and have recently been used for ASD classification. We work on the ASD classification using the LSTM and GRU-based framework in this project.  To understand LSTM and GRU models, first, we must understand the evolution of a Recurrent Neural Network (RNN) from a basic neural network. The structure of the basic neural network can be broken down into three components: input layer, hidden layer, and output layer. The Perceptron is the first trainable neural network and the simplest neural network with only one layer with adjustable weights and thresholds lying between input and output layers. RNN is a type of neural network that deals explicitly with time series or sequential data. RNNs are the improvements of the Feed-Forward Neural Networks (FFNN), where the signals can travel only one way from input to output. FFNNs can only read the current input layer and lacks an internal memory; therefore, these are bad at predicting what is to come. The most widely studied and used feed-forward neural network model is the multilayer FFNN, also called multilayer perceptrons. RNNs are called recurrent because they perform the same task for every element of a sequence, with the output being dependent on the previous computations. RNN is better than FFNN in modeling sequential data because RNN has a memory that captures the information about what has been calculated so far, and the output from the previous steps is taken as an input for the current stage. Hence, RNN makes a prediction, or a decision, based on its current input and the outputs from the previous step.\\
\\
Mathematically speaking, the basic version of RNN has the form  $h_t = g(W x_t + W_{h}h_{t-1}+b) $,
where $x_t$ is the $k$ dimensional input vector at time $t$, $h_t $ the $d-$dimensional hidden state, $g$ is the activation function (such as the logistic function, the hyperbolic tangent function, or the rectified Linear Unit), and $W \in \mathbb{R}^{d \times k}, W_h \in \mathbb{R}^{d \times d}$ are weight matrices, and $b \in \mathbb{R}^{d \times 1}$ is a bias vector~\cite{lipton2015critical}. As we can see, RNN is revolutionary compared to FFNNs, but there are some limitations. In the normal RNN cell, the input at a time step and the hidden state from the previous time step is passed through the activation function to obtain a new hidden state and output where the gradients carry information used in the RNN parameter update. The parameter updates become insignificant when the gradient tends to zero, which means no real learning is done. On the other hand, when the error gradients accumulate during an update, the explosion occurs through exponential growth by repeatedly multiplying gradients through the network layers with values larger than 1, resulting in large gradients. These, in turn, result in large updates to the network weights and an unstable network. Hence due to the vanishing or exploding nature of the (stochastic) gradients with long sequences, RNN has difficulty in learning long-term dependencies \cite{hochreiter1998vanishing, bengio1994learning, pascanu2013difficulty}. For this reason, a different and improved architecture of RNNs was created called, LSTM.\\
\\
LSTM essentially extends the memory of RNNs so that it can handle longer sequences of information. Unlike a standard RNN, which is comprised of the input, hidden, and output layers, LSTM has a memory cell that consists of an input gate, forget gate, and an output gate~\cite{hochreiter1997LSTM, gers1999LSTM, gers2003LSTM, sherstinsky2020fundamentals}. The most crucial component of LSTM architecture is the cell state which runs through the chain, with only linear interaction and keeping the flow of information unchanged. The gate mechanism of LSTM deletes or modifies the information of the cell state. First, the input gate decides which information is received, and then it goes through the forget gate, which determines if it is essential information to keep, and then it makes its way to the output gate. The LSTM utilizes a sigmoid function, a tanh function, and a pointwise multiplication operation to decide which information is passed through based on its importance. 
The architecture of LSTM at time $t$ is presented in the Fig~\ref{Figure:LSTM_architecture}. In this figure, the four gates \textemdash output, change, input, and forget \textemdash are shown with their operations at time $t$.
For a given input sequence $\{ x_1, x_2, \dots, x_n\} $, $ x_t \in \mathbb{R}^{k \times 1}$ is the input sequence at time t. The memory cell $c_t$ updates the information using three gates: input gate $i_t$, forget gate $f_t$, and change gate $\tilde{c_t}$. The hidden state $h_t$ is updated using output gate $o_t$ and the memory cell $c_t$. 
The respective gates and layers compute the following functions at time $t$:
\begin{align*}  
 & i_t = \sigma(W_{i} x_t + W_{hi}h_{t-1}+b_{i}), \\
 & f_t = \sigma(W_{f} x_t + W_{hf}h_{t-1}+b_{f}), \\
 & o_t = \sigma(W_{o} x_t + W_{ho}h_{t-1}+b_{o}), \\
 & \tilde{c_{t}}= \tanh(W_{c} x_t + W_{hc}h_{t-1}+b_{c}), \\
 & c_t = f_t \odot c_{t-1} + i_t \odot \tilde{c_{t}}, \\
 & h_t = o_t \odot \tanh(c_t)
\end{align*} 
where, $\sigma$ represent the sigmoid function. The operator $\odot$ is the element-wise product, $W \in \mathbb{R}^{d \times k}, W_h \in \mathbb{R}^{d \times d}$ are weight matrices, and $b \in \mathbb{R}^{d \times 1}$ are bias vectors. Furthermore, $n, k, d $ denotes the sequence length, the number of features, and the hidden size respectively~\cite{greff_et_al_2017_LSTM, Lei2019_LSTM, dvornek2017identifying, lindemann2021survey}. Several works on the literature during most recent years can be found on the fMRI data classification using LSTM neural network~\cite{khullar2021deep, liu2020multi,  dvornek2019jointly, el2019hybrid}.
In addition, to incorporate the required dimension of LSTM architecture, the input sequence $X_t$ is created by taking $m$ continuous sequence $ x_t: x_{t+m-1}$ which is a matrix of shape $k \times m$ for $t \in \{1, 2, \dots, n-m-1\}$. The output $h_t$ of LSTM is a feature representation for the input sequence $X_t$ at time $t$. Mathematically, $h_t$ can be expressed as follows:
$$ h_t= LSTM(X_t, h_{t-1}, c_{t-1}, w)$$ 
where $w $ denotes all learnable parameters. Since the final hidden state $h_f$ encodes the most information from the input sequence, it is converted to a vector using a dense layer.  
\subsection*{A brief overview of GRU}
\label{Methods_Review_GRU}
\begin{figure}[h]
  \includegraphics[scale=.32]{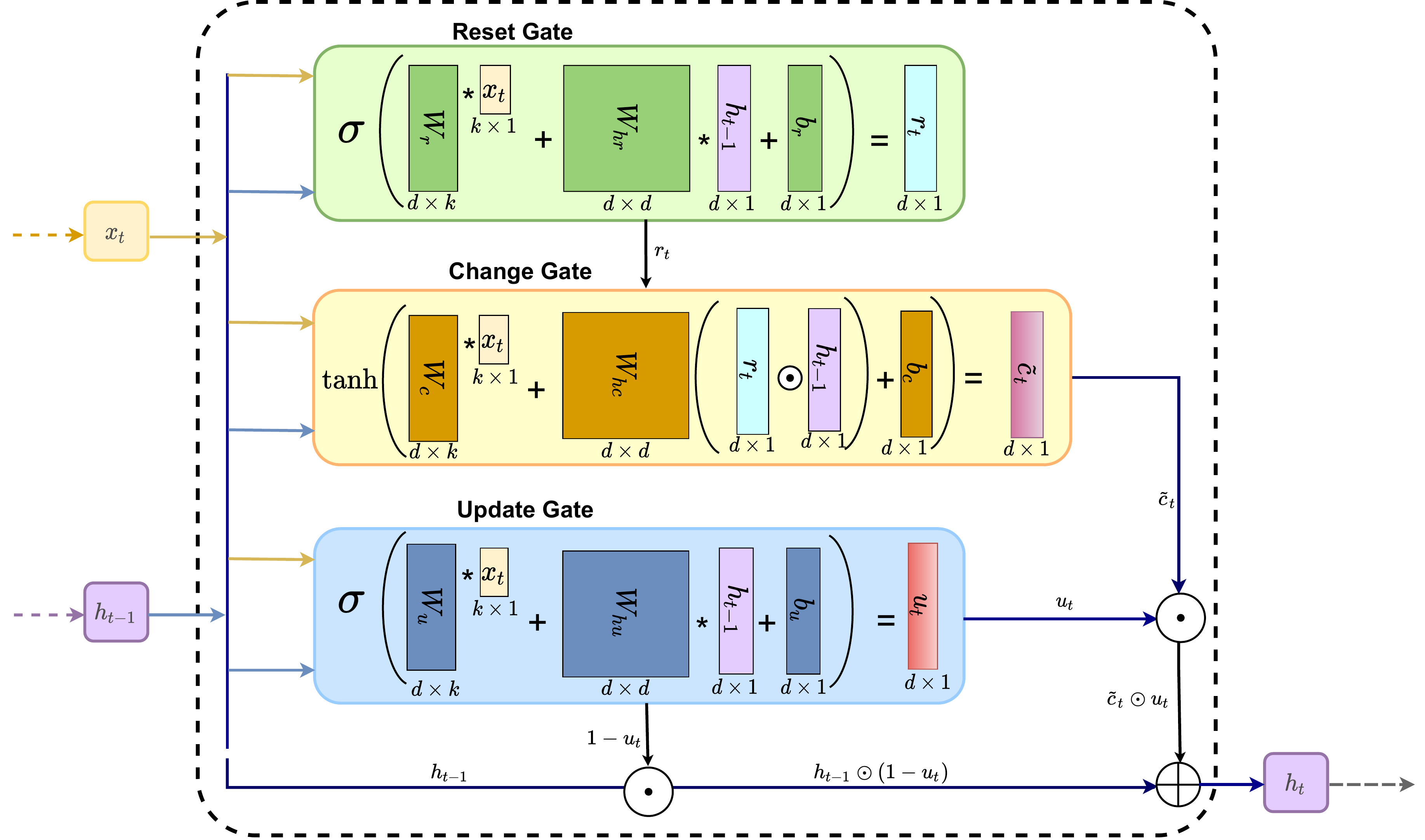}
 \caption{{\bf GRU architecture~\cite{pokhrel2022predicting}.}}
 \label{Figure:GRU_architecture}
\end{figure}

GRU proposed by~\cite{Cho2014LearningPR}, is a relatively recent model inspired by LSTM that can deal with the vanishing or exploding nature of the (stochastic) gradients with long sequences. GRU minimizes or simplifies the three gates from the previous LSTM down to two. The two gates in GRU are called update and reset gates. The update gate is a combination of LSTM's input gate and the forget gate. Both the update gate and reset gate are two vectors that are calculated to determine which information goes through~\cite{dey2017gate, chung2014empirical}.
The reset gate, denoted by $r_t$, is calculated by plugging in $x_t$ and then multiplying by its weight $W_r$ and then added with the multiplication of the previous information denoted by $h_{t-1}$ to its weighs $W_{hr}$. These so-called weighted values, $W_r$ and $W_{hr}$, are matrices. Next, the sigmoid function denoted by $\sigma$, is applied to the sum of the previous results, which will result in a real number that falls between $0$ and $1.$ The reset gate decides how much information from the previous times should not go through. Then a $\tanh$ activation function, a nonlinear function which is denoted by $\tilde{c}_t$, is applied to assist in determining which information should be kept. Next, the update gate, denoted by $u_t$, is calculated similarly to the reset gate. However, the differences are in the weight matrices and the application of the update gate, which decides how much past information should go through. Finally, the final output is calculated by summating the element-wise multiplication to the update gate and the $\tanh$ function. 
The architecture of GRU at time $t$ is presented in Fig~\ref{Figure:GRU_architecture}. In this figure, the three gates \textemdash reset, change and update \textemdash are shown with their operations at time $t$.
Similar as for LSTM model, for a given input sequence $\{ x_1, x_2, \dots, x_n\} $, with $ x_t \in \mathbb{R}^{k \times 1}$ the input sequence at time t. 
The gates and layers for the GRU model  compute the following functions at time $t$:
\begin{align*}  
 & r_t = \sigma(W_{r} x_t + W_{hr}h_{t-1}+b_{r}), \\%reset gate
 & u_t = \sigma(W_{u} x_t + W_{hu}h_{t-1}+b_{u}), \\%update gate
 & \tilde{c_{t}} = \tanh(W_{c} x_t + W_{c} (r_t \odot h_{t-1}) +b_{c}), \\%change gate
 & h_t = (1 - u_t) \odot h_{t-1} + u_t \odot \tilde{c_{t}} %hidden layer at time t or output
\end{align*} 
where, $\sigma$ represent the sigmoid function, $\odot$ is the element-wise product; $W \in \mathbb{R}^{d \times k}, W_h \in \mathbb{R}^{d \times d}$ are weight matrices, and $b \in \mathbb{R}^{d \times 1}$ are bias vectors. In addition, $n, k, d $ denotes the sequence length, the number of features, and the hidden size respectively~\cite{dey2017gate}. 
The input sequence $X_t$ for the GRU model is created  similarly as for the LSTM model. The output $h_t$ of GRU is a feature representation for the input sequence $X_t$ at time $t$, expressed mathematically as follows:
$$ h_t= GRU(X_t, h_{t-1}, c_{t-1}, w)$$ 
where $w $ denotes all parameters that can be learned by training the model. 
%%%%%%%%%%%%%%%%%%%%%%%%%%%%%%%%%%%%%%%

\subsection*{Algorithms}
\label{algorithms}

The pseudo-code of our computational framework, especially the neural network-based GRU and LSTM models, is provided in this section. The algorithms for other machine learning models can be found in the~\cite{suthaharan2016machine}. Here, the algorithm~\ref{GRU_LSTM_Model_Par_Tun} provides the pseudo-code for the hyperparameter tuning procedure for the LSTM and GRU models. Finally, the algorithm~\ref{GRU_LSTM_Model_Alg} shows how the LSTM and GRU models are incorporated in the computational framework for the ASD classification. The LSTM and GRU function were used from the keras-nightly 2.5.0. with the tensorflow version 2.5.0 for the computational work. 
%%%%%%%%%%%%%%%%%%%%%%%

%ALGORITHM VERSION 3

\begin{algorithm}
\caption{\bf Pseudo Code for Hyperparameter Tuning Procedure for GRU/LSTM Model}
\label{GRU_LSTM_Model_Par_Tun}
Input Preparation: Read features after/without PCA transformation; Split train, validation and test data sets; and create input of the form [\#subjects, time step, \#features]\\
Model Input: [\#subjects, time step, \#features]; choices of  optimizers, learning rates, and batch sizes.\\
Initialize: Set the number of epochs sufficiently large.
\begin{tabbing}
(nr)ss\=ijkl\=bbb\=ccc\=ddd\= eeeeee  \kill
\quad{\bf For} ``choice of optimizers'', {\bf Do} \\
\quad\>{\bf For} ``choice of learning rates'', {\bf Do} \\
\quad\>\>{\bf For} ``choice of batch sizes'', {\bf Do} \\
\quad\>\>\> {\bf For} ``range of number of replicates'', {\bf Do} \\
\quad\>\>\>\> Train the model, monitor validation loss;\vspace{.08cm}\\
\quad\>\>\>\> {\bf Continue} Until validation loss at epoch $n \leq$ validation loss at epoch  \\ 
\quad\>\>\>\>\> $n+1 \leq$ validation loss at epoch $n +2$ Or maximum epochs reached.\vspace{.08cm}\\
\quad\>\>\>\> Evaluate model on the validation data.\vspace{.08cm}\\
\quad\>\>\>\>  Calculate accuracy scores.\vspace{.08cm}\\
\quad\>\>\> {\bf End Do}.\vspace{.08cm}\\
\quad\>\>\>  Calculate average accuracy scores.\vspace{.08cm}\\
\quad\>\> {\bf End Do}.\vspace{.08cm}\\
\quad\> {\bf End Do}.\vspace{.08cm}\\
\quad {\bf End Do}.\vspace{.08cm}\\
{\bf Output} Set of best hyperparameters, average accuracy scores, \\
best average accuracy score. 
\end{tabbing}
\end{algorithm}
%%%%%%%%%%%%%%%%%%%%%%%%%%%%%%%%%%%%%%%%%%%%%%%%%%%%%%%%%%%%%%%%%%%%

%%%%%%%%%%%%%%%%%%%%%%%%%%%%%%%%%%%%%%%%%%%%%%%%%%%%%%%%%%%%%%%%%%%%%%
\begin{algorithm}
\caption{\bf Pseudo Code for GRU/LSTM Model}
\label{GRU_LSTM_Model_Alg}
Input Preparation: Read features after/without PCA transformation; Split train and test data sets; and create input of the form [\#subjects, time step, \#features]\\
Model Input: [\#subjects, time step, \#features];  chosen hyperparameters  (optimizer, learning rate, batch size) obtained from Algorithm 1 for each model.\\
Initialize: Set the number of epochs sufficiently large.
\begin{tabbing}
(nr)ss\=ijkl\=bbb\=ccc\=ddd\= eeeeee  \kill
\quad{\bf For} ``choice of neurons'', {\bf Do} \\
\quad\> {\bf For} ``range of number of replicates'', {\bf Do} \\
\quad\>\> Train the model, monitor training loss;\vspace{.08cm}\\
\quad\>\> {\bf Continue} Until validation loss at epoch $n \leq$ validation loss at epoch \\ 
\quad\>\>\> $n+1 \leq$ validation loss at epoch $n +2$ Or maximum epochs reached.\vspace{.08cm}\\
\quad\>\> Evaluate model on the test data.\vspace{.08cm}\\
\quad\>\>  Calculate accuracy, sensitivity, specificity and AUC scores.\vspace{.08cm}\\
\quad\> {\bf End Do}.\vspace{.08cm}\\
\quad\>  Calculate minimum, maximum, average and standard deviation of \\
\quad\> accuracy, sensitivity, specificity and AUC scores.\vspace{.08cm}\\
\quad\> Save the results in respective files.\\
\quad {\bf End Do}.\vspace{.08cm}\\
\end{tabbing}
\end{algorithm}

%%%%%%%%%%%%%%%%%%%%%%%%%%%%%%%%%%%%%%%%%%%%%%%%%%%%%%%%%%%%%%%%%%%%%%%%%%%%

\subsection{Data Description and Preprocessing}
\label{Methods_Data-preprocess}
\begin{figure}[h!]
\centering
  \includegraphics[scale=0.7]{ 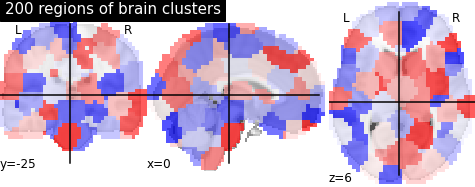}
  \caption{{\bf CC200 atlas by Cameron and Craddock }}%\cite{craddock2012whole}
   \label{Figure:CC200 ROI}
\end{figure}

ABIDE~\cite{craddock2013neuro} initiative has aggregated functional brain imaging data collected from laboratories around the world to accelerate our understanding of the neural bases of autism. ABIDE is an open source for preprocessed neuroimaging data shared by the preprocessed connectomes project. We obtained the resting-state fMRI data from the ABIDE. Once rs-fMRI data are obtained, the first data analysis stage is preprocessing after initial quality control. The primary purpose of preprocessing is to reduce the effects of artifacts and other noise in preparation for FC analysis. The preprocessing of the ABIDE data was done with the X version of the Conjugate Analysis Scalable Pipeline (CPAC), which includes the following: slice time correction, motion correction, temporal filtering, skull stripping, nuisance regression, normalization, and registration. First, functional images were registered to anatomical space by a linear transformation, followed by white matter boundary-based transformation. The white matter boundary-based transformation was accomplished using FMRIB's Linear Image Registration Tool of FMRIB Software Library and white matter tissue segmentation of FAST.
The fMRI resting-state data is now ready for FC analysis after preprocessing. Many different functional atlases have been available such as CC200/CC400, BASC197/444, Power 264 region atlas, to name a few. ROIs generated from functional atlas, such as CC200/CC400, outperform the anatomical atlas (AAL, HO) in the context of resting-state FC analysis \cite{craddock2012whole}, \cite{bellec2010multi}. Since we are focused on analyzing both interpretability and accuracy, the functional connectivity in this paper is calculated based on the functional atlas CC200. Figure \ref{Figure:CC200 ROI} shows the brain ROIs of CC200 atlas. 
A total of 871 subjects (observations) was selected for the analysis. The preprocessed fMRI data for each subject is a matrix of $196 \times 200 $ where the matrix's first dimension(rows) represents the time points and the second dimension(columns) of the matrix depicts the brain regions of interest. The 200 brain regions of interest are obtained based on the functional atlas CC200. Among the 871 observations under consideration, 403 are autistic, and the remaining 468 are normal.
 %
%
%%%%%%%%%%%%%%%%%%%%%%%%%%%%%%%%
\subsection{Generation of Functional Connectivity}
\label{Methods_fun-connectvitiy}
 After extracting the ROIs time series based on the CC200 atlas, the network of FC is created. We use three different methods to construct the connectivity matrix: Pearson's Correlation, Spearman's Rank Correlation, and Partial Correlation. The Pearson's correlation coefficient~ \cite{pearson1895correlation} is chosen as it is the widely used connectivity measure in the literature. On the other hand, partial correlation is claimed to be the better measure of the direct connectivity between the two brain regions stating that the correlation coefficient may seem to be higher; however, the correlation may be because of the influence of the other regions ~[\cite{Beer2019}, \cite{kim2015testing}]. In addition, Spearman's rank correlation~\cite{artusi2002bravais} is used since it measures both linear and nonlinear relationships between the brain regions of interest. The connectivity matrices were symmetric ($m\ \times \ m$), where the upper diagonal elements were extracted into a feature vector of length $\frac{m\left(m-1\right)}{2}$. The elements on the diagonals were ignored as they represent the connections with themselves. Each feature represents the correlation between two regions of the brain conditioned on other regions. Here, we have data of the size $871 \times 19,900$ with the number of rows representing the number of subjects and the number of columns representing the features.
%
%
%%%%%%%%%%%%%%%%%%%%%%%%%%%%%%%%
\subsection{Dimension Reduction using Principal Component Analysis (PCA)}
\label{Methods_dim_red}
\begin{figure}[h!]
  \centering
  \subfloat[]{\includegraphics[width=0.35\textwidth]{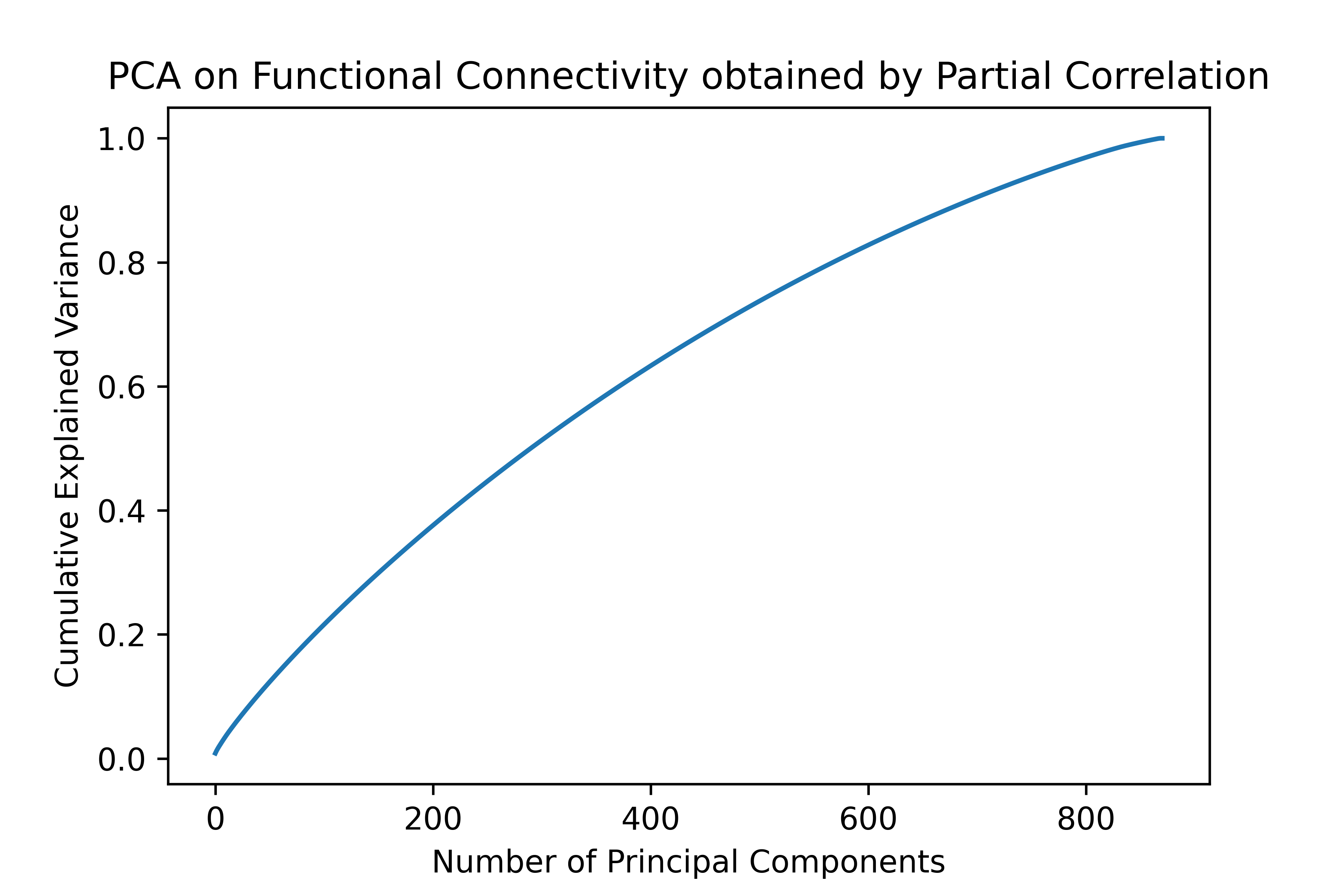}} 
  \subfloat[]{\includegraphics[width=0.35\textwidth]{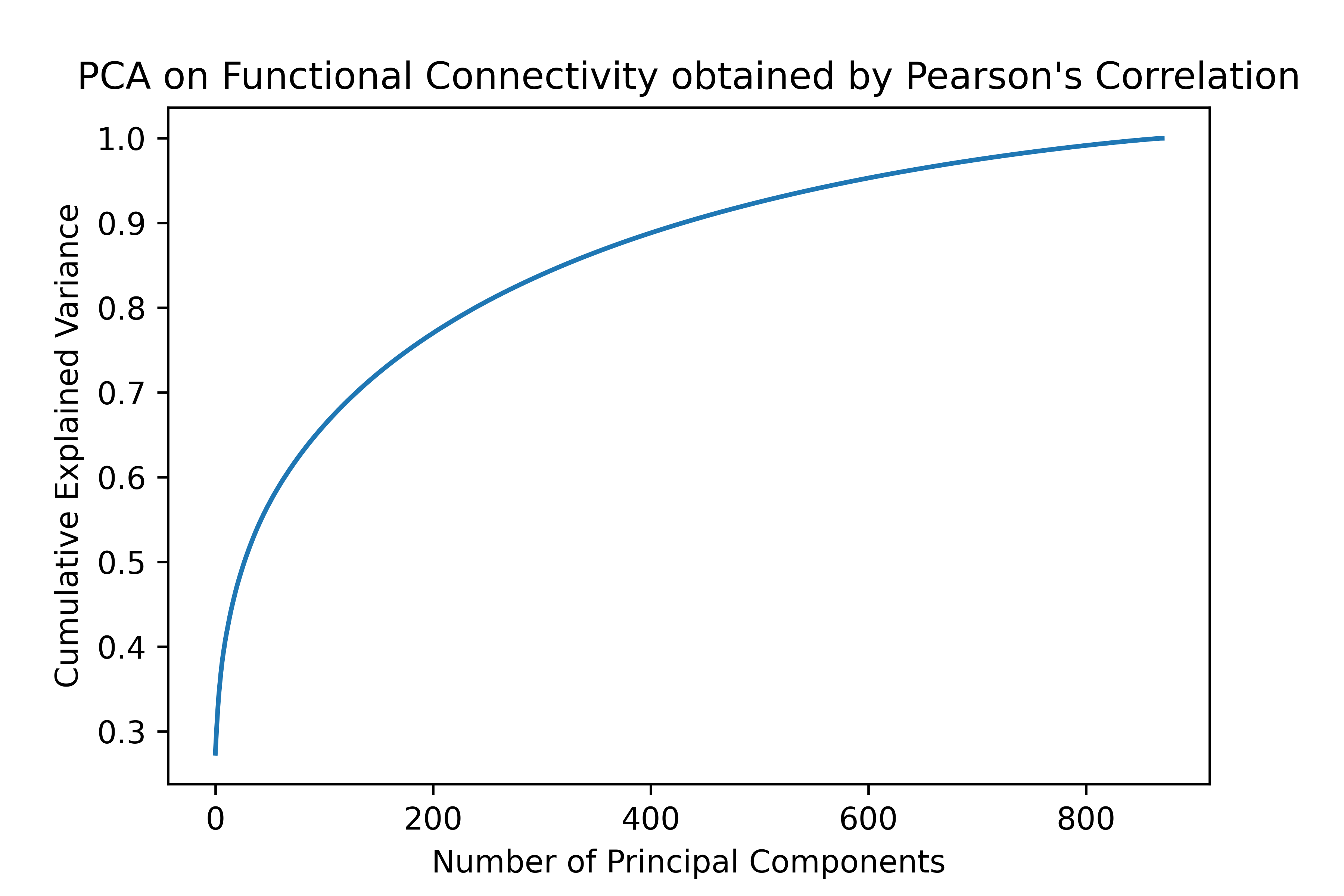}} 
  \subfloat[]{\includegraphics[width=0.35\textwidth]{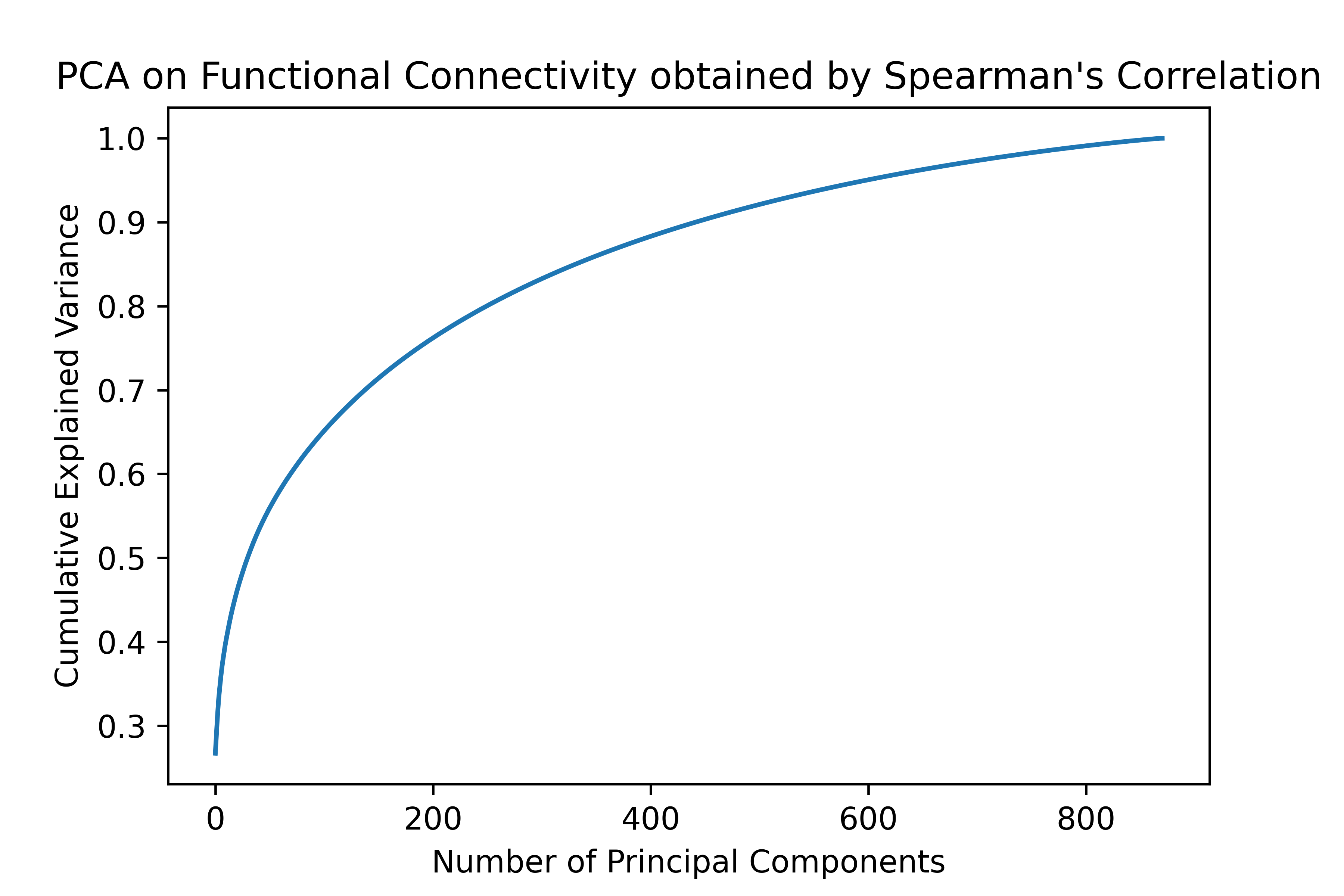}}
  \caption{The percentage of the variance explained by the number of principal components on the features obtained by (a)Partial Correlation,  (b) Pearson's Correlation, and (c) Spearman's Correlation.}
  \label{fig:DimRed_PCA}
\end{figure}

We have 871 observations with 19,900 features on each of them. Hence, we encounter the high-dimensional statistics-related problem where the number of features is much higher than the sample size. The PCA is implemented to mitigate the common challenges with high-dimensional data. For example, computational expense and an increased error rate due to multiple test corrections when testing each feature for association with an outcome. PCA simplifies the complexity in high-dimensional data by transforming the data into fewer dimensions which acts as summaries of features~\cite{lever2017points},\cite{altman2018curse}. After the PCA was implemented, around 80 percent of the data variability can be explained by 600 principal components. Figure~\ref {fig:DimRed_PCA} shows the percentage of variance explained by the principal components for all three different RSFC. %871 600-dimensional vectors
%%%%%%%%%%%%%%%%%%%%%%%%%%%%%%%%%%
%
%
%%%%%%%%%%%%%%%%%%%%%%%%%%%%%%%%%%
The 600 principal components that correspond to each type of FC were chosen. Each subject then has 600 features, each representing the principal components, and we have observations on the 871 subjects for the analysis. In this case, our data is of size $871 \times 600$ with the number of rows representing the number of subjects and the number of columns representing the number of principal components chosen. The experiment is performed on two types of data, one without PCA and another after applying PCA.
%
%
%
%
%%%%%%%%%%%%%%%%%%%%%%%%%%%%%%%%%%
\section{Experiments and Results} 
%%%%%%%%%%%%%%%%%%%%%%%%%%%%%%%%%%%%%%
\subsection{Experimental Methods}
\label{experimental_methods}
Several supervised learning methods such as LR, LSVC, KSVC, RFC, ABC, GRU, and LSTM neural networks are experimented in this article for ASD classification.  The scikit-learn library~\cite{scikit-learn}, \cite{bhandari2022lstm} and the TensorFlow~\cite{tensorflow2015-whitepaper} are mainly used for the implementation of the methods in python. The input for the models is the features vector obtained using three different methods for constructing FC. The experiment is performed on two different types of data; features extracted from the original data with PCA and features extracted from the original data without PCA. First, ten-fold cross-validation is performed on the whole data to develop the classification model. Then, the best value of the parameters is selected from the model. Finally, the repeated ten-fold cross-validation is performed with these parameters, and the performance on the hold-out fold is reported. For stable performance, the experiment is replicated ten times. In addition, the feature selection method called the recursive feature selection technique was implemented. This technique takes all the features and drops the least important feature each time until the desired number of features to select is reached. In our analysis, the number of features is reduced to half of the total features whenever the recursive feature elimination method is implemented. The same procedure discussed above is repeated on the model with the application of recursive feature elimination and the model without implementing the recursive feature elimination method. \\

To fit the LSTM and GRU model, the necessary steps of the data preparation have been taken to prepare the input data as described below.
The architecture for the LSTM neural network is presented in Figure~\ref{Figure:LSTM_architecture}. First, the input sequence is created using $k=19,900$ features without dimension reduction ( k = 600 after dimension reduction using PCA) with time step 1, as shown in the top part of the figure.
Then at time $t-1$, the input $X_{t-1}$, a matrix of size $k \times 1$ , together with $h_{t-2}$ and $c_{t-2}$ is fed into the LSTM. For the next step, the output $h_{t-1}$ of the previous step, together with input sequence $X_t$ and the cell memory $c_{t-1}$ become input for LSTM. This process continues until the final input sequence $X_f$ with corresponding output $h_f$, a vector of length equal to the given number of neurons of the last LSTM layer. Finally, $h_f$ is transmitted to a fully connected layer where the sigmoid function is used to predict the class of chess masters and chess novices.
The transformed data at this point is a two-dimensional array (number of observations, number of features). However, the LSTM model expects three-dimensional input (number of observations, time step, number of predictors). Therefore, the time step of 1 is chosen to make the input data compatible with the model.  Finally, the data is split randomly into two parts: 80\% of the data is allocated for the training set, and the remaining 20\% of the data is for the test set.  The data preparation steps for GRU model is exactly same as of LSTM model. 

For the LSTM and GRU models, the single-layer models have been implemented. For the models with various neurons ranges from 10 to 250, the hyperparameters to choose for each models are \textemdash  the learning rate, the optimizers, the batch sizes,  and the number of epochs. The various learning rates between 0.1 to 0.0001, the number of neurons between 10 to 250, the batch sizes in the range of  4 to 32, and the optimizers Adam, Adamax, and Nadam from Keras were experimented with. Once the best value of the hyperparameters is chosen, the model is fit on the training data using the hyperparameters chosen, and the performance is evaluated on the test data set. Both LSTM and GRU models are also replicated ten times for stable model performance.
%
%%%%%%%%%%%%%%%%%%%%%%%%%%%%%%%%%%
\subsection{Hyperparameter Tuning}
\label{Hyperparameter Tuning}
For the LR, LSVC, KSVC, ABC, and RFC models, the ten-fold cross-validation method was used to find the best value of the hyperparameters. LR was implemented with $L_1$, $L_2$, and elasticnet penalties. The feature selection techniques did not improve the results on the lasso penalty implemented version. The best result from the LR model was obtained using the feature selection method known as 'SelectFromModel' with a ridge penalty on logistic regression. For the KSVC model, the radial basis function kernel was used. For the RFC, the maximum depth of the tree between 5 and 20, the number of trees between 500 and 2500, and the node impurity criterion gini and entropy were compared, and the model with 2500 trees and the maximum depth of 5 were fitted using gini coefficient to minimize the node impurity. The AdaBoost classifier was implemented with a learning rate of 0.1, 0.01, and 0.001 and a number of trees between 500 and 1500. The model with a learning rate of 0.01 and 500 estimators was selected to build the final model. 

For the LSTM and GRU model, the data set was divided into three parts, a training set, a validation set, and a test set. Initially, the data is split 80 percent into a training set and the remaining 20 percent into a test set. Then the 20\% data from the training set is used for the validation set. The hyperparameters for the LSTM and GRU model are the batch sizes, the learning rate, the optimizers, and the number of epochs. We let the epochs be 100 and apply the early stopping criterion from the Keras library; the model will stop training if there is no improvement in the validation loss five consecutive times. Since the value of epochs is allowed to be sufficiently high, the early stopping criterion chooses the appropriate value of epochs. The remaining hyperparameters were chosen using hyperparameter tuning. The choice of the optimizers is Adam, Adagrad, and Nadam. The learning rate is chosen among the values of 0.01, 0.05, 0.001, 0.005, and 0.0001. The batch sizes of 4, 8, 16, and 32 were considered. The models with the number of neurons 10, 30, 50, 100, 150, 200, and 250 were compared. Overall, we have seven models for both LSTM and GRU neural networks. There are 5*4*3 = 60 combinations for each of them, and the best combination among them is chosen using validation data. Since we have three sets of data obtained from Spearman RSFC, Pearson RSFCs, and Partial RSFCs, we performed the hyperparameter tuning for every data model. The best combination of hyperparameters was chosen that corresponds to the highest accuracy on the validation data. The chosen hyperparameters values were used to build the final model. 
%%%%%%%%%%%%%%%%%%%%%%%%%%%%%%%%%%%%%%%%%%%%%%%%%%%%%%%%
%\textcolor{red}{Table number 1}
\begin{table}[h!]
\begin{center}
	\caption{
	{\bf Hyperparameter tuning for 10 neurons single layer LSTM on the Pearson's RSFC.}}
	\label{Table:hyperparameter tuning 10 neurons single layer LSTM on the Pearson's RSFC}
		%\resizebox{13cm}{!}{
\begin{tabular}{|c|c|c|c|c|c|}
\hline
\multirow{2}{*}{Optimizer} & \multirow{2}{*}{Learning rate} & \multicolumn{4}{c|}{Batch size} \\ \cline{3-6} 
                           &                                & 4         & 8        & 16    & 32   \\ \hline
\multirow{5}{*}{Adam}      & 0.01             & 0.6086     & 0.6136    & 0.6145 & \bf{0.6163}  \\ \cline{2-6} 
                           & 0.005                        & 0.6126      & 0.6115    & 0.6116 & 0.6098  \\ \cline{2-6} 
                            & 0.001                       & 0.6090      & 0.6102    & 0.6099 & 0.6094  \\ \cline{2-6}
                           & 0.0005                      & 0.6093      & 0.6093    & 0.6094 & 0.6091  \\ \cline{2-6} 
                           & 0.0001                      & 0.6088      & 0.6091    & 0.6079 & 0.6063   \\ \hline%%%%%%%%%
\multirow{5}{*}{Nadam}     & 0.01            & 0.6070    & 0.6081  & 0.6081  &  0.6079  \\ \cline{2-6} 
			 & 0.005               & 0.6075    & 0.6077  & 0.6074  & 0.6068  \\ \cline{2-6} 
                           & 0.001                        & 0.6052    & 0.6027   & 0.6004  & 0.5988   \\ \cline{2-6} 
                           & 0.0005                      & 0.5971    & 0.5947   & 0.5924  & 0.5909 \\ \cline{2-6} 
                           & 0.0001                      & 0.5890    & 0.5870    & 0.5847  & 0.5828    \\ \hline%%%%%%%
\multirow{5}{*}{Adagrad}   & 0.01            & 0.5838    & 0.5841   & 0.5844   & 0.5848  \\ \cline{2-6} 
			& 0.005                & 0.5855    & 0.5859   & 0.5861   & 0.5867 \\ \cline{2-6} 
                           & 0.001                        & 0.5874    & 0.5877   & 0.5878   & 0.5883  \\ \cline{2-6} 
                           & 0.0005                      & 0.5887    & 0.5893   & 0.5895   & 0.5900   \\ \cline{2-6} 
                           & 0.0001                      & 0.5903    & 0.5903   & 0.5903  &  0.5894 \\ \hline
\end{tabular}
\end{center}
\end{table}
%
%%%%%%%%%%%%%%%%%%%%%%%%%%%%%%%%%%%%%%%%%%%%%%%%%%%%%%%%%%%%%%
%
For example, the hyperparameter tuning results for the 10 neurons single layer LSTM model is given in the Table~\ref{Table:hyperparameter tuning 10 neurons single layer LSTM on the Pearson's RSFC}. The best accuracy on the validation data was obtained with the batch size of 32 with a learning rate of 0.01 on the Adam optimizer, so these values of hyperparameters were chosen for the ten neurons single layer LSTM model. The hyperparameter tuning results for all other models are available in the appendix from Table~\ref{Table:Hyperparameter tuning for 30 neurons single layer LSTM on the Pearson's RSFC} - Table~\ref{Table:Hyperparameter tuning for 250 neurons single layer GRU with Partial RSFC}.
Finally, the model fits the training data using the chosen hyperparameters, and the performance is evaluated on the test data set. For a stable model performance, each model is replicated ten times.
%

%%%%%%%%%%%%%%%%%%%%%%%%%%%%%%%%%%%%%%%%%%%%%%%%%%%%%%%%%%%%%%%%%%%
\subsection{Model Performance}
\label{Methods_model_performance}
The developed framework is tested on classifying ASD patients with healthy patients. For the LR, SVC, KSVC, ABC, and RFC models, the selected value of parameters obtained from the grid-search cross-validation has been used in the repeated ten-fold cross-validation. The ten-fold cross-validation is replicated ten times, and the average accuracy, sensitivity, specificity, and AUC scores obtained from the hold-out fold are reported. In addition, the average performance scores from these methods implementing the recursive feature selection and the select from model techniques are recorded. The feature selection technique gives a mixed result with a smaller improvement in the accuracy of the LR model in comparison to without using model selection. The overall best performance is obtained from the logistic regression model with the feature selection. The best score was the accuracy of 71\%, the sensitivity of 68\%, specificity of 73\%, and area under the curve score of 77\%. The results from the LSTM and GRU models are close to the best results obtained.\\
\\
For the LSTM and GRU models, various single-layer models of 10, 30, 50, 100, 150, 200, and 250 neurons have been built with the best values of the hyperparameters chosen. The model is trained using the training data and tested on the 20\% unseen data separated before the training. The early stopping criterion is implemented during the model training to stop training if there is no improvement in the loss for three consecutive times. We compare the results of the 21 models( seven models with three different RSFCs) for each of the LSTM and GRU. Each model is replicated ten times, and the average performance scores were reported for better reproducibility. Below, we report the best results for features obtained from each RSFC in a separate table.
The following tables summarize the performance of all implemented methods obtained by using the Pearson's correlation, Partial correlation, and Spearman's rank correlation RSFC methods.\\
\\
%
%%%%%%%%%%%%%%%%%%%%%%%%%%%%%%%%%%%%%%%%%%%%%%%%%%%%%%%%%%%%%%%%%%%
%\textcolor{red}{Table number 3: Generic style}
\begin{table}[h!]
	\caption{
	{\bf Performance scores of the various supervised learning models on Partial RSFC using PCA}}
	\label{Table:ResultsPar_RSFC}
	%\resizebox{13cm}{!}{
\begin{center}
\begin{tabular}{|c|c|c|c|c|c|c|c|c|}
\hline
Metrices                      & Methods  & KSVC & LSVC & LR +SFM & RFC & ABC & LSTM & GRU \\ \hline
\multirow{4}{*}{Accuracy}         & Min             & 47.13    & 43.68    & 43.18    & 47.12 & 51.15   & 58.86    & 59.43    \\ \cline{2-9} 
                              & Average         & 58.61 & 56.11 &  { \bf 59.66} & 56.18 & 56.74 & 59.14 & { \bf 60.69}     \\ \cline{2-9} 
                              & Max             & 66.67    & 68.97    & 68.18   & 67.82 &  63.22  &  60.00   & 62.86    \\ \cline{2-9} 
                              & Std             & 4.46    & 5.06    & 3.62    & 4.56  & 3.36  & 0.49     & 1.14    \\ \hline
\multirow{4}{*}{Sensitivity}         & Min             & 30    &32.50    & 15.00   & 25 &  26.25  & 55.13    & 55.41     \\ \cline{2-9} 
                              & Average         & 45.27 &  53.94 & 29.75   &40.76 & 40.17  &  56.67 & 57.57  \\ \cline{2-9} 
                              & Max             & 60.98    &70    & 47.50    & 60 &  53.09 &  57.69   & 66.34    \\ \cline{2-9} 
                              & Std             & 6.79    &7.50    & 6.24    & 6.76  &  5.99 &   0.81  & 1.31    \\ \hline
\multirow{4}{*}{Specificity}            & Min             &53.19    &42.55    &74.47  &   56.52 & 59.57  & 59.79    &   60.40  \\ \cline{2-9} 
                              & Average        & 70.08 &  58 &  85.41 & 69.45 &  71 &  61.13 & 62.97  \\ \cline{2-9} 
                              & Max             & 85.11    & 72.34    & 95.74    & 85.11 &  81.91  &  61.86   & 66.34    \\ \cline{2-9} 
                              & Std             & 6.70    & 6.45    & 5.12   & 6.42  & 5.77  &  0.70   &  1.88   \\ \hline
\multirow{4}{*}{AUC}            & Min             & 48.30    &42.50    & 52.01    &  42.71  & 50.63 & 58.46     &59.60     \\ \cline{2-9} 
                              & Average       & 61.52 &  56.69 & 64.36 & 57.55  &57.45  &  58.78 & 60.61     \\ \cline{2-9} 
                              & Max             & 74.02    & 71.49    & 78.47    & 71.85   & 66.36   & 59.25    & 61.37     \\ \cline{2-9} 
                              & Std             & 5.22    & 5.42    & 5.12    & 5.60   & 3.39   &  0.25   &   0.49  \\ \hline
\end{tabular}
\end{center}
\end{table}
%%%%%%%%%%%%%%%%%%%%%%%%%%%%%%%%%%%%%%%%%%%%%%%%%%%%%%%%%%%%%%
Table~\ref{Table:ResultsPar_RSFC} summarizes the performance scores of the various supervised learning models on Partial RSFC using PCA. All results were average model evaluation scores calculated based on ten replications on the test dataset. The minimum, maximum, mean, and standard deviation scores were reported for each metric. The accuracy scores achieved by all the tested models on Partial RSFC using PCA ranges, on average, between 56.11\%–60.69\%. The proposed GRU exhibited a significantly higher accuracy $(60.69\% \pm 1.14\%)$ than KSVC, LSVC, LR+SFM, RFC, ABC, and LSTM models. The sensitivity scores obtained from KSVC, LSVC, LR+SFM, RFC, ABC, LSTM and GRU on the dataset were 45.27\%, 53.94\%, 29.75\%, 40.76\%, 40.17\%, 56.67\% and 57.57\%, respectively. Consistent with the accuracy results, GRU also had better sensitivity $(57.57\%\pm 1.31\%)$ than all other compared models. While the specificity $(85.41\% \pm 5.12\%)$ and AUC $(64.36\% \pm 5.12\%)$ of the LR+SFM were significantly higher than GRU, we noted that the LR+SFM had a higher standard deviation simultaneously, suggesting that the results had a wider spread as compared to other methods. Moreover, the LR+SFM method had a lower sensitivity $(29.75\% \pm 6.24\%)$. Therefore, LR+SFM did not achieve the best mean precision in our experiments. We also observed that the LSTM had the lowest standard deviation, among other approaches. However, the performance of LSTM was inferior to GRU in terms of accuracy, sensitivity, and AUC rates. Given the proposed GRU achieved an improved classification accuracy $(60.69\%\pm 1.14\%)$, sensitivity $(57.57\% \pm 1.31\%)$, specificity $(62.97\%\pm 1.88\%)$, and AUC $(60.61\%\pm 1.88\%)$ among compared machine learning models, and therefore the GRU had the highest precision, and the LSTM is the closest competitor.\\
\\
%
%
%%%%%%%%%%%%%%%%%%%%%%%%%%%%%%%%%%%%%%%%%%%%%%%%%%%%%%%%%%%%%%%%%%%
%\textcolor{red}{Table number 3: Generic style}
\begin{table}[h!]
	\caption{
	{\bf Performance scores of the various supervised learning models on Pearson's RSFC using PCA}}
	\label{Table:ResultsPear_RSFC}
	%\resizebox{13cm}{!}{
\begin{center}
\begin{tabular}{|c|c|c|c|c|c|c|c|c|}
\hline
Metrices                      & Methods  & KSVC & LSVC & LR +SFM & RFC & ABC & LSTM & GRU \\ \hline
\multirow{4}{*}{Accuracy}         & Min             & 56.32    &47.13    & 60.92   & 47.73 & 54.02  & 64.00    & 67.16    \\ \cline{2-9} 
                              & Average         & 67.27 &  62.08 & { \bf 71.17} &  62.04 & 60.10 & 67.42 & { \bf 69.16}  \\ \cline{2-9} 
                              & Max             & 79.31    & 73.56   & 86.36   & 78.16 & 68.57    & 70.29    & 71.37    \\ \cline{2-9} 
                              & Std             & 4.92    &4.79    & 4.92   & 4.87 & 3.19  & 2.19    &  1.45   \\ \hline
\multirow{4}{*}{Sensitivity}         & Min             &40    &42.50    & 52.50    & 19.51 &  30.86  &  56.41   & 66.37     \\ \cline{2-9} 
                              & Average         & 60.77 &  60.47 &  68.88 &  37.46 & 41.27 &  62.69 & 69.47   \\ \cline{2-9} 
                              & Max             & 77.50   &77.50    & 85.37   & 57.50    & 54.30 &  67.95   &  71.68   \\ \cline{2-9} 
                              & Std             & 8.53    &7.70    &7.81    & 8.43   & 5.36 & 4.11    & 2.26    \\ \hline
\multirow{4}{*}{Specificity}            & Min             & 59.57    & 38.30    &61.70    & 67.39 & 66.67   & 67.01     & 67.11    \\ \cline{2-9} 
                              & Average        & 72.86 &  63.46 &  73.14 &  82.22 & 76.32 &  71.24 & 68.93  \\ \cline{2-9} 
                              & Max             & 85.11    & 78.26   &89.36    & 95.74    & 89.25 & 75.26    & 71.14    \\ \cline{2-9} 
                              & Std             & 5.78    &7.09    &5.68    & 5.38   & 4.87 & 2.89    &  1.34   \\ \hline
\multirow{4}{*}{AUC}            & Min             & 58.51    &53.67    & 66.06    & 55   & 53.98 & 69.39     & 73.09    \\ \cline{2-9} 
                              & Average       & 74.07 & 66.99 &  77.38 &  66.88 & 63.53 & 72.35 & 74.15   \\ \cline{2-9} 
                              & Max             & 84.10    &77.36    & 87.40   & 81.06  &72.75   &75.48     & 75.04     \\ \cline{2-9} 
                              & Std             & 4.97   &4.67    &4.51    & 5.37     & 3.64 & 1.86    &   0.72  \\ \hline
\end{tabular}
\end{center}
\end{table}
%%%%%%%%%%%%%%%%%%%%%%%%%%%%%%%%%%%%%%%%%%%%%%%%%%%%%%%%%%%%%%
%
 %
Table~\ref{Table:ResultsPear_RSFC} presents the average performance scores of the various supervised learning models on Pearson's RSFC using PCA. To assess the performance of various supervised learning models dealing with the ASD classification problem on Pearson's RSFC,  the experiments using  KSVC, LSVC, LR+SFM, RFC, ABC, LSTM, and GRU, have been conducted.  The final evaluations were determined concerning each approach's average accuracy, sensitivity, specificity, and AUC metrics. The two best performing approaches were LR+SFM and GRU with the classification accuracies of $71.17\% \pm 4.92\%$ and $69.16\% \pm 1.45\%$, respectively. While the sensitivity of GRU $(69.47\% \pm 2.26\%)$ was only slightly greater than LR+SFM $(68.88\% \pm7.81\%)$, the lower standard deviation signified its highly stable performance on detecting ASD patients in comparison to LR+SFM. Despite RFC having the highest specificity $(82.22\% \pm5.38\%)$, it also obtained, among other methods, the lowest sensitivity $(37.46\% \pm8.43\%)$, namely, it successfully identified the normal controls but overlooked actual ASD subjects significantly. LR+SFM was superior to GRU by achieving a better AUC score of $77.38\% \pm 4.51\%$. However, LR+SFM had higher standard deviations of the accuracy, sensitivity, specificity, and AUC scores, demonstrating its weak classification consistency. In general, the best performing method achieved an overall accuracy, sensitivity, specificity, and AUC scores of $69.16\% \pm 1.45\%, 69.47\% \pm2.26\%, 69.47\% \pm2.26\%$ and $74.15\% \pm0.72\%$, using GRU approach based on Pearson's RSFC, and the LR+SFM is the closest competitor.\\
%
%%%%%%%%%%%%%%%%%%%%%%%%%%%%%%%%%%%%%%%%%%%%%%%%%%%%%%%%%%%%%%%%%%%%
%\textcolor{red}{Table number 3: Generic style}
\begin{table}[h!]
	\caption{
	{\bf Performance scores of the various supervised learning models on Spearman's RSFC using PCA }}
	\label{Table:ResultsSpear_RSFC}
	%\resizebox{13cm}{!}{
\begin{center}
\begin{tabular}{|c|c|c|c|c|c|c|c|c|}
\hline
Metrices                      & Methods  & KSVC & LSVC & LR +SFM & RFC & ABC & LSTM & GRU \\ \hline
\multirow{4}{*}{Accuracy}         & Min             &56.32    & 48.28    &56.32    & 51.72 &  53.14  & 64.57    & 66.03    \\ \cline{2-9} 
                              & Average         & 67.41 & 61.83 &  { \bf 68.48} &  61.48 & 60.69 & 66.17 & { \bf 68.28}  \\ \cline{2-9} 
                              & Max             & 79.31    &72.41    & 78.16   & 70.11   & 67.82  & 70.29    &  70.61 \\ \cline{2-9} 
                              & Std             & 5.06    &4.72    &4.14    & 3.87    & 2.95 & 1.74    &  1.58   \\ \hline
\multirow{4}{*}{Sensitivity}         & Min             & 40    & 45    & 48.78    & 25   & 34.57 & 58.97    &  64.23   \\ \cline{2-9} 
                              & Average         & 60.97 &  59.69 &  65.58 &  40.66 & 45.21 & 62.69 & 67.80  \\ \cline{2-9} 
                              & Max             & 80.49    &77.50    & 82.93    & 55   & 55.56 & 69.23    & 71.22    \\ \cline{2-9} 
                              & Std             & 7.70    &7.07    &7.29    & 6.35    & 4.97 & 3.22     &1.88     \\ \hline
\multirow{4}{*}{Specificity}            & Min             & 56.52    & 46.81    &53.19    & 65.22 &63.44   & 65.98    &  65.47   \\ \cline{2-9} 
                              & Average        & 72.94 &  63.66 &  70.98 &  79.40 & 74 &  68.99 & 68.71 \\ \cline{2-9} 
                              & Max             & 86.96    & 80.85    &82.98    & 89.36    & 84.04 &  73.20   &  71.22   \\ \cline{2-9} 
                              & Std             & 6.88    & 7.13   & 5.71   & 5.15    & 4.71 & 2.09   & 1.83    \\ \hline
\multirow{4}{*}{AUC}            & Min             & 61.33    & 48.14    & 64.20    & 55.32    & 58.05 & 70.94    & 73.05    \\ \cline{2-9} 
                              & Average       & 73.55 & 65.88 &  74.88 &  67.39 &64.69 & 72.07 & 74.49    \\ \cline{2-9} 
                              & Max             & 85.11    & 77.45    & 89.50   &  77.50   & 72.63 & 73.45    &76.95     \\ \cline{2-9} 
                              & Std             & 4.95    & 5.44   &4.67    &  4.77   & 3.34 &0.79     &1.17     \\ \hline
\end{tabular}
\end{center}
\end{table}
%%%%%%%%%%%%%%%%%%%%%%%%%%%%%%%%%%%%%%%%%%%%%%%%%%%%%%%%%%%%%%
%
To assess how our model behaves on Spearman's RSFC using PCA, we observed the accuracy, sensitivity, specificity, and AUC metrics on the unseen data for each supervised learning model experimented. Table~\ref{Table:ResultsSpear_RSFC} presents the average performance scores of all the supervised learning models implemented on Spearman's RSFC using PCA. In Table~\ref{Table:ResultsSpear_RSFC}, we observed that the LR+SFM achieved a mean classification accuracy of $68.48\% \pm 4.14\%$ (sensitivity $65.58\% \pm 7.29\%$, specificity $70.98\% \pm 5.17\%$, and AUC $74.88\% \pm 4.67\%$), obtained from ten replications on the test dataset. While the RFC approach achieved a relatively low average accuracy of $61.48\% \pm 3.87\%$ (sensitivity $40.66\% \pm 6.35\%$, specificity $79.40\% \pm 5.15\%$, and AUC $67.39\% \pm 4.77\%$), it attained the highest specificity, illustrating its outstanding capability in accurately identifying control group. Meanwhile, the GRU classifier obtained a mean accuracy of $68.28\% \pm 1.58\%$ (sensitivity $67.80\% \pm 1.88\%$, specificity $68.71\% \pm 1.83\%$, and AUC $74.49\% \pm 1.17\%$). The result showed that the sensitivity of GRU was superior to the other six classifiers on the dataset; namely, GRU had the highest ability to designate an individual with the disease as positive, and thus fewer cases of the disease were missed. In addition, GRU had lower standard deviations in accuracy, sensitivity, specificity, and AUC, demonstrating its strong classification consistency. Thus, based upon the results of Spearman's RSFC, GRU was selected to generate classification systems that have better performance in terms of sensitivity, specificity, and accuracy among all the compared models.\\

From tables~\ref{Table:ResultsPar_RSFC}, \ref{Table:ResultsPear_RSFC}, and~\ref{Table:ResultsSpear_RSFC}, we observed that the average performance scores of the LR + SFM model on the Pearson’s RSFC using PCA is higher among all other compared methods. On the other hand, the popular deep learning methods GRU and LSTM have the closest scores with a lower variation. From the application point of view, the GRU model is more reliable and chosen as the best model, and the LSTM and LR + SFM model are the closest competitors. The results have some limitations from the inference perspective as the features are selected by applying PCA on the RSFCs. The most important brain regions to identify the ASD cannot be accurately determined as the features are the principal components which is the linear combination of the 19,900 original features.
%
%%%%%%%%%%%%%%%%%%%%%%%%%%%%%
\section{Discussion and Conclusion}
\label{discussion}
%%%%%%%%%%%%%%%%%%%%%%%%%%%%%
%
In this article, we classified autistic and normal patients using various supervised learning methods. The RSFC obtained from Pearson's correlation, Spearman's rank correlation, and the Partial correlation was used. The performance of the classification models corresponding to each rs-RSFC measure is compared using the accuracy, specificity, sensitivity, and roc AUC scores. Since the total number of observations is 871 and the number of features is 19,900, our model faces the curse of dimensionality, so the PCA was implemented for dimension reduction. The experiments are conducted on the original data and the data after implementing PCA. The application of PCA improves the model performance and significantly reduces the computational time at the cost of interpretability of features. The supervised learning methods LR, KSVC, LSVC, RFC, ABC, LSTM, and GRU, were implemented on the original and transformed data with PCA. LSTM and GRU model with various neurons value from 10 to 250 with a range of values of the hyperparameters is compared, and the simplest model with the best accuracy is chosen. Pearson's RSFC is considered a better measure of connectivity for the ASD classification as the highest performance scores are obtained on PCA-transformed Pearson's RSFC data. Although the LR model has overall higher average performance scores, this model has a high variance. The GRU model has slightly lower accuracy, specificity, and AUC scores and slightly higher sensitivity scores. However, the GRU model has a much smaller variance. So the GRU model is considered the best model and is chosen for the classification. The LSTM model remains competitive in model accuracy and sensitivity and has a consistent performance measure compared to other methods. For precision, our experiments suggest the GRU method to be preferable. The classification result on the original data is much worse with high computational cost in comparison to the PCA transformed data.\\
\\
%Our results suggest that the simpler models LR and SVC are preferable to the more complex machine learning models LSTM and GRU for this problem. 
%
%
Our work has some limitations. The proposed model frameworks are less beneficial for the statistical inference point of view. Since the principal components are the linear combination of the original variables with corresponding weight values, it is difficult to identify which brain regions contribute more to classifying autistic and normal patients. Without the implementation of PCA, these methods can be utilized if inference is our main objective. In that case, we see a significant loss in precision. Also, as the sample size is small, the overall performance scores may depend on the split of the data set. In addition, the LSTM neural network expects to have a decent sample size to learn the value of the parameters. Therefore, the model performance may be improved with a larger data set. Exploring the hybrid classification techniques, implementing various global optimizers and local optimizers in the model, applying other dimension reduction techniques such as independent component analysis, autoencoders, and manifold learning methods is a matter of future work.

\bmhead{Acknowledgments}

This research work is partially supported by MTSU FRCAC grant.\\

%\section*{Declarations}
%
%Some journals require declarations to be submitted in a standardised format. Please check the Instructions for Authors of the journal to which you are submitting to see if you need to complete this section. If yes, your manuscript must contain the following sections under the heading `Declarations':
%
%\begin{itemize}
%\item Funding
%\item Conflict of interest/Competing interests (check journal-specific guidelines for which heading to use)
%\item Ethics approval 
%\item Consent to participate
%\item Consent for publication
%\item Availability of data and materials
%\item Code availability 
%\item Authors' contributions
%\end{itemize}
%%===========================================================================================%%
\begin{appendices}
\subsection{ \large\bf Hyperparameter tuning for single layer LSTM on various RSFCs}
\label{ Hyperparameter tuning for single layer LSTM on various RSFCs}
\subsection*{ Hyperparameter tuning for single layer LSTM on the Pearson's RSFC}
\label{ Hyperparameter tuning for single layer LSTM on the Pearson's RSFC}
\begin{table}[h!]
\begin{center}
	\caption{
	{\bf Hyperparameter tuning for 30 neurons single layer LSTM on the Pearson's RSFC.}}
	\label{Table:Hyperparameter tuning for 30 neurons single layer LSTM on the Pearson's RSFC}
		%\resizebox{13cm}{!}{
\begin{tabular}{|c|c|c|c|c|c|}
\hline
\multirow{2}{*}{Optimizer} & \multirow{2}{*}{Learning rate} & \multicolumn{4}{c|}{Batch size} \\ \cline{3-6} 
                           &                                & 4         & 8        & 16    & 32   \\ \hline
\multirow{5}{*}{Adam}      & 0.01             & 0.6179     & 0.6204    & 0.6174 & 0.6143  \\ \cline{2-6} 
                           & 0.005                        & 0.6121      & 0.6117    & 0.6104 & 0.6118  \\ \cline{2-6} 
                            & 0.001                       & 0.6111      & 0.6115    & 0.6123 & 0.6123  \\ \cline{2-6}
                           & 0.0005                      & 0.6126      & 0.6127    & 0.6130 & 0.6132  \\ \cline{2-6} 
                           & 0.0001                      & 0.6132      & 0.6129    & 0.6130 & 0.6121   \\ \hline%%%%%%%%%
\multirow{5}{*}{Nadam}     & 0.01            & 0.6117    & 0.6127  & 0.6130  &  0.6131  \\ \cline{2-6} 
			 & 0.005               & 0.6137    & 0.6142  & 0.6151  & 0.6154  \\ \cline{2-6} 
                           & 0.001                        & 0.6137    & 0.6119   & 0.6096  & 0.6076   \\ \cline{2-6} 
                           & 0.0005                      & 0.6054    & 0.6033   & 0.6010  & 0.5988 \\ \cline{2-6} 
                           & 0.0001                      & 0.5967    & 0.5943    & 0.5920  & 0.5897    \\ \hline%%%%%%%
\multirow{5}{*}{Adagrad}   & 0.01            & 0.5900    & 0.5905   & 0.5910   & 0.5913  \\ \cline{2-6} 
			& 0.005                & 0.5920    & 0.5924   & 0.5927   & 0.5928 \\ \cline{2-6} 
                           & 0.001                        & 0.5934    & 0.5938   & 0.5938   & 0.5941  \\ \cline{2-6} 
                           & 0.0005                      & 0.5944    & 0.5949   & 0.5950   & 0.5954   \\ \cline{2-6} 
                           & 0.0001                      & 0.5954    & 0.5956   & 0.5960  &  0.5966 \\ \hline
\end{tabular}
\end{center}
\end{table}
%\vspace{-2 in}
%%%%%%%%%%%%%%%%%%%%%%%%%%%%%%%%%%%%%%%%%%%%%%%%%%%%%%%%%%%%%%
%3)	Hyperparameter tuning for 50 neurons single layer LSTM on the Pearson's RSFC.
%\newpage
%%%%%%%%%%%%%%%%%%%%%%%%%%%%%%%%%%%%%%%%%%%%%%%%%%%%%%%%
%\textcolor{red}{Table number 2}
\begin{table}[h!]
\begin{center}
	\caption{
	{\bf Hyperparameter tuning for 50 neurons single layer LSTM on the Pearson's RSFC.}}
	\label{Table:Hyperparameter tuning for 50 neurons single layer LSTM on the Pearson's RSFC.}
		%\resizebox{13cm}{!}{
\begin{tabular}{|c|c|c|c|c|c|}
\hline
\multirow{2}{*}{Optimizer} & \multirow{2}{*}{Learning rate} & \multicolumn{4}{c|}{Batch size} \\ \cline{3-6} 
                           &                                & 4         & 8        & 16    & 32   \\ \hline
\multirow{5}{*}{Adam}      & 0.01             & 0.6286     & 0.6232    & 0.6210 & 0.6225  \\ \cline{2-6} 
                           & 0.005                        & 0.6216      & 0.6207    & 0.6193 & 0.6190  \\ \cline{2-6} 
                            & 0.001                       & 0.6186      & 0.6184    & 0.6171 & 0.6154  \\ \cline{2-6}
                           & 0.0005                      & 0.6163      & 0.6154    & 0.6162 & 0.6163  \\ \cline{2-6} 
                           & 0.0001                      & 0.6176      & 0.6166    & 0.6166 & 0.6156   \\ \hline%%%%%%%%%
\multirow{5}{*}{Nadam}     & 0.01            & 0.6159    & 0.6160  & 0.6171  &  0.6176  \\ \cline{2-6} 
			 & 0.005               & 0.6183    & 0.6187  & 0.6188  & 0.6190  \\ \cline{2-6} 
                           & 0.001                        & 0.6180    & 0.6171   & 0.6165  & 0.6141   \\ \cline{2-6} 
                           & 0.0005                      & 0.6124    & 0.6109   & 0.6085  & 0.6064 \\ \cline{2-6} 
                           & 0.0001                      & 0.6034    & 0.6006    & 0.5981  & 0.5955    \\ \hline%%%%%%%
\multirow{5}{*}{Adagrad}   & 0.01            & 0.5961    & 0.5966   & 0.5970   & 0.5972  \\ \cline{2-6} 
			& 0.005                & 0.5975    & 0.5978   & 0.5979   & 0.5982 \\ \cline{2-6} 
                           & 0.001                        & 0.5986    & 0.5990   & 0.5993   & 0.5997  \\ \cline{2-6} 
                           & 0.0005                      & 0.5600    & 0.6002   & 0.6007   & 0.6011   \\ \cline{2-6} 
                           & 0.0001                      & 0.6014    & 0.6016   & 0.6017   & 0.6020 \\ \hline
\end{tabular}
\end{center}
\end{table}

%%%%%%%%%%%%%%%%%%%%%%%%%%%%%%%%%%%%%%%%%%%%%%%%%%%%%%%%%%%%%%
%4)	Hyperparameter tuning for 100 neurons single layer LSTM on the Pearson's RSFC.
%\newpage
%%%%%%%%%%%%%%%%%%%%%%%%%%%%%%%%%%%%%%%%%%%%%%%%%%%%%%%%
%\textcolor{red}{Table number 2}
\begin{table}[h!]
\begin{center}
	\caption{
	{\bf Hyperparameter tuning for 100 neurons single layer LSTM on the Pearson's RSFC.}}
	\label{Table:Hyperparameter tuning for 100 neurons single layer LSTM on the Pearson's RSFC}
		%\resizebox{13cm}{!}{
\begin{tabular}{|c|c|c|c|c|c|}
\hline
\multirow{2}{*}{Optimizer} & \multirow{2}{*}{Learning rate} & \multicolumn{4}{c|}{Batch size} \\ \cline{3-6} 
                           &                                & 4         & 8        & 16    & 32   \\ \hline
\multirow{5}{*}{Adam}      & 0.01             & 0.6129     & 0.6118    & 0.6139 & 0.6123  \\ \cline{2-6} 
                           & 0.005                        & 0.6126      & 0.6131    & 0.6139 & 0.6142  \\ \cline{2-6} 
                            & 0.001                       & 0.6144      & 0.6145    & 0.6151 & 0.6146  \\ \cline{2-6}
                           & 0.0005                      & 0.6152      & 0.6158    & 0.6166 & 0.6168  \\ \cline{2-6} 
                           & 0.0001                      & 0.6161      & 0.6163    & 0.6165 & 0.6164   \\ \hline%%%%%%%%%
\multirow{5}{*}{Nadam}     & 0.01            & 0.6174    & 0.6181  & 0.6182  &  0.6186  \\ \cline{2-6} 
			 & 0.005               & 0.6194    & 0.6199  & 0.6208  & 0.6208  \\ \cline{2-6} 
                           & 0.001                        & 0.6208    & 0.6201   & 0.6187  & 0.6173   \\ \cline{2-6} 
                           & 0.0005                      & 0.6168    & 0.6160   & 0.6143  & 0.6123 \\ \cline{2-6} 
                           & 0.0001                      & 0.6093    & 0.6065    & 0.6038  & 0.6014    \\ \hline%%%%%%%
\multirow{5}{*}{Adagrad}   & 0.01            & 0.6015    & 0.6019   & 0.6022   & 0.6023  \\ \cline{2-6} 
			& 0.005                & 0.6027    & 0.6029   & 0.6032   & 0.6036 \\ \cline{2-6} 
                           & 0.001                        & 0.6036    & 0.6038   & 0.6042   & 0.6045  \\ \cline{2-6} 
                           & 0.0005                      & 0.6044    & 0.6048   & 0.6049   & 0.6051   \\ \cline{2-6} 
                           & 0.0001                      & 0.6056    & 0.6058   & 0.6060   & 0.6062 \\ \hline
\end{tabular}
\end{center}
\end{table}
%\vspace{-2 in}
%%%%%%%%%%%%%%%%%%%%%%%%%%%%%%%%%%%%%%%%%%%%%%%%%%%%%%%%%%%%%%
%\newpage
%5)	Hyperparameter tuning for 150 neurons single layer LSTM on the Pearson's RSFC.

%%%%%%%%%%%%%%%%%%%%%%%%%%%%%%%%%%%%%%%%%%%%%%%%%%%%%%%%
%\textcolor{red}{Table number 2}
\begin{table}[h!]
\begin{center}
	\caption{
	{\bf Hyperparameter tuning for 150 neurons single layer LSTM on the Pearson's RSFC.}}
	\label{Table:Hyperparameter tuning for 150 neurons single layer LSTM on the Pearson's RSFC}
		%\resizebox{13cm}{!}{
\begin{tabular}{|c|c|c|c|c|c|}
\hline
\multirow{2}{*}{Optimizer} & \multirow{2}{*}{Learning rate} & \multicolumn{4}{c|}{Batch size} \\ \cline{3-6} 
                           &                                & 4         & 8        & 16    & 32   \\ \hline
\multirow{5}{*}{Adam}      & 0.01             & 0.6121     & 0.6164    & 0.6167 & 0.6123  \\ \cline{2-6} 
                           & 0.005                        & 0.6147      & 0.6151    & 0.6153 & 0.6163  \\ \cline{2-6} 
                            & 0.001                       & 0.6167      & 0.6172    & 0.6173 & 0.6176  \\ \cline{2-6}
                           & 0.0005                      & 0.6187      & 0.6185    & 0.6185 & 0.6184  \\ \cline{2-6} 
                           & 0.0001                      & 0.6185      & 0.6183    & 0.6177 & 0.6179   \\ \hline%%%%%%%%%
\multirow{5}{*}{Nadam}     & 0.01            & 0.6180    & 0.6186  & 0.6195  &  0.6199  \\ \cline{2-6} 
			 & 0.005               & 0.6201    & 0.6204  & 0.6207  & 0.6212  \\ \cline{2-6} 
                           & 0.001                        & 0.6213    & 0.6218   & 0.6210  & 0.6195   \\ \cline{2-6} 
                           & 0.0005                      & 0.6187    & 0.6176   & 0.6164  & 0.6144 \\ \cline{2-6} 
                           & 0.0001                      & 0.6123    & 0.6101   & 0.6075  & 0.6049    \\ \hline%%%%%%%
\multirow{5}{*}{Adagrad}   & 0.01            & 0.6049    & 0.6052   & 0.6053   & 0.6056  \\ \cline{2-6} 
			& 0.005                & 0.6059    & 0.6060   & 0.6062   & 0.6064 \\ \cline{2-6} 
                           & 0.001                        & 0.6066    & 0.6067   & 0.6070   & 0.6071  \\ \cline{2-6} 
                           & 0.0005                      & 0.6072    & 0.6073   & 0.6076   & 0.6083   \\ \cline{2-6} 
                           & 0.0001                      & 0.6084    & 0.6085   & 0.6086   & 0.6089 \\ \hline
\end{tabular}
\end{center}
\end{table}
%\vspace{-2 in}
%%%%%%%%%%%%%%%%%%%%%%%%%%%%%%%%%%%%%%%%%%%%%%%%%%%%%%%%%%%%%%
%6)	Hyperparameter tuning for 200 neurons single layer LSTM on the Pearson's RSFC.
%\newpage
%%%%%%%%%%%%%%%%%%%%%%%%%%%%%%%%%%%%%%%%%%%%%%%%%%%%%%%%
%\textcolor{red}{Table number 2}
\begin{table}[h!]
\begin{center}
	\caption{
	{\bf Hyperparameter tuning for 200 neurons single layer LSTM on the Pearson's RSFC.}}
	\label{Table:Hyperparameter tuning for 200 neurons single layer LSTM on the Pearson's RSFC}
		%\resizebox{13cm}{!}{
\begin{tabular}{|c|c|c|c|c|c|}
\hline
\multirow{2}{*}{Optimizer} & \multirow{2}{*}{Learning rate} & \multicolumn{4}{c|}{Batch size} \\ \cline{3-6} 
                           &                                & 4         & 8        & 16    & 32   \\ \hline
\multirow{5}{*}{Adam}      & 0.01             & 0.6121     & 0.6236    & 0.6240 & 0.6230  \\ \cline{2-6} 
                           & 0.005                        & 0.6234      & 0.6223    & 0.6233 & 0.6251  \\ \cline{2-6} 
                            & 0.001                       & 0.6252      & 0.6244    & 0.6256 & 0.6267  \\ \cline{2-6}
                           & 0.0005                      & 0.6274      & 0.6283    & 0.6295 & 0.6292  \\ \cline{2-6} 
                           & 0.0001                      & 0.6301      & 0.6305    & 0.6318 & 0.6323   \\ \hline%%%%%%%%%
\multirow{5}{*}{Nadam}     & 0.01            & 0.6327    & 0.6337  & 0.6347  &  0.6351  \\ \cline{2-6} 
			 & 0.005               & 0.6356    & 0.6362  & 0.6371  & 0.6378  \\ \cline{2-6} 
                           & 0.001                        & 0.6383    & 0.6393   & 0.6389  & 0.6377   \\ \cline{2-6} 
                           & 0.0005                      & 0.6375    & 0.6366   & 0.6356  & 0.6332 \\ \cline{2-6} 
                           & 0.0001                      & 0.6300    & 0.6273   & 0.6244  & 0.6221    \\ \hline%%%%%%%
\multirow{5}{*}{Adagrad}   & 0.01            & 0.6218    & 0.6219   & 0.6219   & 0.6219  \\ \cline{2-6} 
			& 0.005                & 0.6218    & 0.6220   & 0.6222   & 0.6222 \\ \cline{2-6} 
                           & 0.001                        & 0.6223    & 0.6227   & 0.6230   & 0.6231  \\ \cline{2-6} 
                           & 0.0005                      & 0.6232    & 0.6234   & 0.6235   & 0.6238   \\ \cline{2-6} 
                           & 0.0001                      & 0.6240    & 0.6245   & 0.6246   & 0.6251 \\ \hline
\end{tabular}
\end{center}
\end{table}
%\vspace{-2 in}
%%%%%%%%%%%%%%%%%%%%%%%%%%%%%%%%%%%%%%%%%%%%%%%%%%%%%%%%%%%%%%
%7)	Hyperparameter tuning for 250 neurons single layer LSTM on the Pearson's RSFC.
%\newpage
%%%%%%%%%%%%%%%%%%%%%%%%%%%%%%%%%%%%%%%%%%%%%%%%%%%%%%%%
%\textcolor{red}{Table number 2}
\begin{table}[h!]
\begin{center}
	\caption{
	{\bf Hyperparameter tuning for 250 neurons single layer LSTM on the Pearson's RSFC.}}
	\label{Table:Hyperparameter tuning for 250 neurons single layer LSTM on the Pearson's RSFC}
		%\resizebox{13cm}{!}{
\begin{tabular}{|c|c|c|c|c|c|}
\hline
\multirow{2}{*}{Optimizer} & \multirow{2}{*}{Learning rate} & \multicolumn{4}{c|}{Batch size} \\ \cline{3-6} 
                           &                                & 4         & 8        & 16    & 32   \\ \hline
\multirow{5}{*}{Adam}      & 0.01             & 0.6271     & 0.6254    & 0.6264 & 0.6264  \\ \cline{2-6} 
                           & 0.005                        & 0.6260      & 0.6265    & 0.6263 & 0.6265  \\ \cline{2-6} 
                            & 0.001                       & 0.6261      & 0.6262    & 0.6260 & 0.6270  \\ \cline{2-6}
                           & 0.0005                      & 0.6276      & 0.6283    & 0.6280 & 0.6283  \\ \cline{2-6} 
                           & 0.0001                      & 0.6299      & 0.6300    & 0.6307 & 0.6311   \\ \hline%%%%%%%%%
\multirow{5}{*}{Nadam}     & 0.01            & 0.6325    & 0.6339  & 0.6348  &  0.6354  \\ \cline{2-6} 
			 & 0.005               & 0.6362    & 0.6373  & 0.6380  & 0.6384  \\ \cline{2-6} 
                           & 0.001                        & 0.6390    & 0.6394   & 0.6397  & 0.6382   \\ \cline{2-6} 
                           & 0.0005                      & 0.6384    & 0.6376   & 0.6366  & 0.6346 \\ \cline{2-6} 
                           & 0.0001                      & 0.6312    & 0.6284   & 0.6266  & 0.6233    \\ \hline%%%%%%%
\multirow{5}{*}{Adagrad}   & 0.01            & 0.6229    & 0.6224   & 0.6224   & 0.6221  \\ \cline{2-6} 
			& 0.005                & 0.6219    & 0.6220   & 0.6219   & 0.6219 \\ \cline{2-6} 
                           & 0.001                        & 0.6219    & 0.6219   & 0.6222   & 0.6225  \\ \cline{2-6} 
                           & 0.0005                      & 0.6228    & 0.6231   & 0.6233   & 0.6233   \\ \cline{2-6} 
                           & 0.0001                      & 0.6237    & 0.6241   & 0.6245   & 0.6248 \\ \hline
\end{tabular}
\end{center}
\end{table}
%\vspace{-1 in}
%%%%%%%%%%%%%%%%%%%%%%%%%%%%%%%%%%%%%%%%%%%%%%%%%%%%%%%%%%%%%%
\clearpage
\subsection*{Hyperparameter tuning for single layer LSTM on the Partial Correlation RSFC}
\label{Hyperparameter tuning for single layer LSTM on the Partial Correlation RSFC}
%Hyperparameter tuning for 10 neurons single layer LSTM on the Partial Corr RSFC.

%%%%%%%%%%%%%%%%%%%%%%%%%%%%%%%%%%%%%%%%%%%%%%%%%%%%%%%%
%\textcolor{red}{Table number 1}
\begin{table}[h!]
\begin{center}
	\caption{
	{\bf Hyperparameter tuning for 10 neurons single layer LSTM on the Partial Corr RSFC.}}
	\label{Table:Hyperparameter tuning for 10 neurons single layer LSTM on the Partial Corr RSFC}
		%\resizebox{13cm}{!}{
\begin{tabular}{|c|c|c|c|c|c|}
\hline
\multirow{2}{*}{Optimizer} & \multirow{2}{*}{Learning rate} & \multicolumn{4}{c|}{Batch size} \\ \cline{3-6} 
                           &                                & 4         & 8        & 16    & 32   \\ \hline
\multirow{5}{*}{Adam}      & 0.01          & 0.5421    & 0.5475    & 0.5514    & 0.5536   \\ \cline{2-6} 
                           & 0.005                     & 0.5537    & 0.5538    & 0.5540    & 0.5543    \\ \cline{2-6} 
                            & 0.001                    & 0.5553    & 0.5556    & 0.5555    & 0.5558    \\ \cline{2-6}
                           & 0.0005                   & 0.5571    & 0.5573    & 0.5566    & 0.5547    \\ \cline{2-6} 
                           & 0.0001                   & 0.5546    & 0.5540    & 0.5538    & 0.5521    \\ \hline%%%%%%%%%
\multirow{5}{*}{Nadam}     & 0.01       & 0.5519     & 0.5518   & 0.5518   &  0.5512     \\ \cline{2-6} 
			 & 0.005         & 0.5523    & 0.5516    & 0.5507   & 0.5487      \\ \cline{2-6} 
                           & 0.001                    & 0.5475    & 0.5457    & 0.5444   & 0.5424      \\ \cline{2-6} 
                           & 0.0005                  & 0.5410    & 0.5404    & 0.5393   & 0.5378      \\ \cline{2-6} 
                           & 0.0001                  & 0.5369    & 0.5361    & 0.5350   & 0.5342      \\ \hline%%%%%%%
\multirow{5}{*}{Adagrad}   & 0.01        & 0.5347    & 0.5349    & 0.5353    & 0.5356    \\ \cline{2-6} 
			& 0.005           & 0.5359   & 0.5366    & 0.5368    & 0.5373     \\ \cline{2-6} 
                           & 0.001                     & 0.5380   & 0.5388    & 0.5393    & 0.5392     \\ \cline{2-6} 
                           & 0.0005                   & 0.5396   & 0.5398    & 0.5398    & 0.5399     \\ \cline{2-6} 
                           & 0.0001                   & 0.5399   & 0.5401    & 0.5402    & 0.5402     \\ \hline
\end{tabular}
\end{center}
\end{table}
%\vspace{-1 in}
%%%%%%%%%%%%%%%%%%%%%%%%%%%%%%%%%%%%%%%%%%%%%%%%%%%%%%%%%%%%%
%\newpage

%Hyperparameter tuning for 30 neurons single layer LSTM on the Partial Corr RSFC.

%%%%%%%%%%%%%%%%%%%%%%%%%%%%%%%%%%%%%%%%%%%%%%%%%%%%%%%%
%\textcolor{red}{Table number 1}
\begin{table}[h!]
\begin{center}
	\caption{
	{\bf Hyperparameter tuning for 30 neurons single layer LSTM on the Partial Corr RSFC.}}
	\label{Table:Hyperparameter tuning for 30 neurons single layer LSTM on the Partial Corr RSFC}
		%\resizebox{13cm}{!}{
\begin{tabular}{|c|c|c|c|c|c|}
\hline
\multirow{2}{*}{Optimizer} & \multirow{2}{*}{Learning rate} & \multicolumn{4}{c|}{Batch size} \\ \cline{3-6} 
                           &                                & 4         & 8        & 16    & 32   \\ \hline
\multirow{5}{*}{Adam}      & 0.01          & 0.5478    & 0.5568    & 0.5550    & 0.5539   \\ \cline{2-6} 
                           & 0.005                     & 0.5523    & 0.5527    & 0.5543    & 0.5537    \\ \cline{2-6} 
                            & 0.001                    & 0.5551    & 0.5567    & 0.5560    & 0.5574    \\ \cline{2-6}
                           & 0.0005                   & 0.5581    & 0.5590    & 0.5583    & 0.5570    \\ \cline{2-6} 
                           & 0.0001                   & 0.5575    & 0.5577    & 0.5575    & 0.5569    \\ \hline%%%%%%%%%
\multirow{5}{*}{Nadam}     & 0.01       & 0.5569     & 0.5572   & 0.5571   &  0.5567     \\ \cline{2-6} 
			 & 0.005         & 0.5576    & 0.5570    & 0.5558   & 0.5544      \\ \cline{2-6} 
                           & 0.001                    & 0.5530   & 0.5516    & 0.5501   & 0.5484      \\ \cline{2-6} 
                           & 0.0005                  & 0.5473    & 0.5457    & 0.5444   & 0.5431      \\ \cline{2-6} 
                           & 0.0001                  & 0.5417    & 0.5406    & 0.5395   & 0.5382      \\ \hline%%%%%%%
\multirow{5}{*}{Adagrad}   & 0.01        & 0.5383    & 0.5388    & 0.5392    & 0.5393    \\ \cline{2-6} 
			& 0.005           & 0.5397   & 0.5399    & 0.5403    & 0.5407     \\ \cline{2-6} 
                           & 0.001                     & 0.5414   & 0.5420    & 0.5424    & 0.5427     \\ \cline{2-6} 
                           & 0.0005                   & 0.5433   & 0.5438    & 0.5441    & 0.5442    \\ \cline{2-6} 
                           & 0.0001                   & 0.5446   & 0.5448    & 0.5447    & 0.5446     \\ \hline
\end{tabular}
\end{center}
\end{table}
%\vspace{-1 in}
%%%%%%%%%%%%%%%%%%%%%%%%%%%%%%%%%%%%%%%%%%%%%%%%%%%%%%%%%%%%%

%\newpage
%Hyperparameter tuning for 50 neurons single layer LSTM on the Partial Corr RSFC.

%%%%%%%%%%%%%%%%%%%%%%%%%%%%%%%%%%%%%%%%%%%%%%%%%%%%%%%%
%\textcolor{red}{Table number 1}
\begin{table}[h!]
\begin{center}
	\caption{
	{\bf Hyperparameter tuning for 50 neurons single layer LSTM on the Partial Corr RSFC.}}
	\label{Table:Hyperparameter tuning for 50 neurons single layer LSTM on the Partial Corr RSFC}
		%\resizebox{13cm}{!}{
\begin{tabular}{|c|c|c|c|c|c|}
\hline
\multirow{2}{*}{Optimizer} & \multirow{2}{*}{Learning rate} & \multicolumn{4}{c|}{Batch size} \\ \cline{3-6} 
                           &                                & 4         & 8        & 16    & 32   \\ \hline
\multirow{5}{*}{Adam}      & 0.01          & 0.5643    & 0.5575    & 0.5536    & 0.5543   \\ \cline{2-6} 
                           & 0.005                     & 0.5520    & 0.5542    & 0.5540    & 0.5542    \\ \cline{2-6} 
                            & 0.001                    & 0.5552    & 0.5572    & 0.5579    & 0.5585    \\ \cline{2-6}
                           & 0.0005                   & 0.5597    & 0.5612    & 0.5611    & 0.5600    \\ \cline{2-6} 
                           & 0.0001                   & 0.5601    & 0.5607    & 0.5602    & 0.5589    \\ \hline%%%%%%%%%
\multirow{5}{*}{Nadam}     & 0.01       & 0.5586     & 0.5588   & 0.5590   &  0.5585     \\ \cline{2-6} 
			 & 0.005         & 0.5588    & 0.5587    & 0.5580   & 0.5556      \\ \cline{2-6} 
                           & 0.001                    & 0.5544   & 0.5527    & 0.5514   & 0.5499      \\ \cline{2-6} 
                           & 0.0005                  & 0.5486    & 0.5475    & 0.5457   & 0.5442      \\ \cline{2-6} 
                           & 0.0001                  & 0.5427    & 0.5419    & 0.5408   & 0.5396      \\ \hline%%%%%%%
\multirow{5}{*}{Adagrad}   & 0.01        & 0.5400    & 0.5404    & 0.5405    & 0.5406    \\ \cline{2-6} 
			& 0.005           & 0.5408   & 0.5411    & 0.5415    & 0.5418     \\ \cline{2-6} 
                           & 0.001                     & 0.5424   & 0.5431    & 0.5434    & 0.5438    \\ \cline{2-6} 
                           & 0.0005                   & 0.5444   & 0.5451    & 0.5455    & 0.5456    \\ \cline{2-6} 
                           & 0.0001                   & 0.5462   & 0.5465    & 0.5466    & 0.5465     \\ \hline
\end{tabular}
\end{center}
\end{table}
%\vspace{-1 in}
%%%%%%%%%%%%%%%%%%%%%%%%%%%%%%%%%%%%%%%%%%%%%%%%%%%%%%%%%%%%%
%\newpage
%Hyperparameter tuning for 100 neurons single layer LSTM on the Partial Corr RSFC.

%%%%%%%%%%%%%%%%%%%%%%%%%%%%%%%%%%%%%%%%%%%%%%%%%%%%%%%%
%\textcolor{red}{Table number 1}
\begin{table}[h!]
\begin{center}
	\caption{
	{\bf Hyperparameter tuning for 100 neurons single layer LSTM on the Partial Corr RSFC.}}
	\label{Table:Hyperparameter tuning for 100 neurons single layer LSTM on the Partial Corr RSFC}
		%\resizebox{13cm}{!}{
\begin{tabular}{|c|c|c|c|c|c|}
\hline
\multirow{2}{*}{Optimizer} & \multirow{2}{*}{Learning rate} & \multicolumn{4}{c|}{Batch size} \\ \cline{3-6} 
                           &                                & 4         & 8        & 16    & 32   \\ \hline
\multirow{5}{*}{Adam}      & 0.01          & 0.5486    & 0.5493    & 0.5543    & 0.5546   \\ \cline{2-6} 
                           & 0.005                     & 0.5567    & 0.5562    & 0.5561    & 0.5559    \\ \cline{2-6} 
                            & 0.001                    & 0.5563    & 0.5574    & 0.5581    & 0.5579    \\ \cline{2-6}
                           & 0.0005                   & 0.5592    & 0.5607    & 0.5613    & 0.5615    \\ \cline{2-6} 
                           & 0.0001                   & 0.5618    & 0.5625    & 0.5631    & 0.5626    \\ \hline%%%%%%%%%
\multirow{5}{*}{Nadam}     & 0.01       & 0.5632     & 0.5632   & 0.5628   &  0.5623     \\ \cline{2-6} 
			 & 0.005         & 0.5627    & 0.5621    & 0.5614   & 0.5600      \\ \cline{2-6} 
                           & 0.001                    & 0.5581   & 0.5574    & 0.5554   & 0.5541      \\ \cline{2-6} 
                           & 0.0005                  & 0.5533    & 0.5523    & 0.5513   & 0.5502      \\ \cline{2-6} 
                           & 0.0001                  & 0.5490    & 0.5479    & 0.5466   & 0.5452      \\ \hline%%%%%%%
\multirow{5}{*}{Adagrad}   & 0.01        & 0.5453    & 0.5459    & 0.5458    & 0.5457    \\ \cline{2-6} 
			& 0.005           & 0.5462   & 0.5465    & 0.5464    & 0.5465     \\ \cline{2-6} 
                           & 0.001                     & 0.5468   & 0.5473    & 0.5476    & 0.5479    \\ \cline{2-6} 
                           & 0.0005                   & 0.5483   & 0.5488    & 0.5492    & 0.5493    \\ \cline{2-6} 
                           & 0.0001                   & 0.5497   & 0.5501    & 0.5502    & 0.5501     \\ \hline
\end{tabular}
\end{center}
\end{table}
%\vspace{-2 in}
%%%%%%%%%%%%%%%%%%%%%%%%%%%%%%%%%%%%%%%%%%%%%%%%%%%%%%%%%%%%%%
%\newpage

%Hyperparameter tuning for 150 neurons single layer LSTM on the Partial Corr RSFC.

%%%%%%%%%%%%%%%%%%%%%%%%%%%%%%%%%%%%%%%%%%%%%%%%%%%%%%%%
%\textcolor{red}{Table number 1}
\begin{table}[h!]
\begin{center}
	\caption{
	{\bf Hyperparameter tuning for 150 neurons single layer LSTM on the Partial Corr RSFC.}}
	\label{Table:Hyperparameter tuning for 150 neurons single layer LSTM on the Partial Corr RSFC}
		%\resizebox{13cm}{!}{
\begin{tabular}{|c|c|c|c|c|c|}
\hline
\multirow{2}{*}{Optimizer} & \multirow{2}{*}{Learning rate} & \multicolumn{4}{c|}{Batch size} \\ \cline{3-6} 
                           &                                & 4         & 8        & 16    & 32   \\ \hline
\multirow{5}{*}{Adam}      & 0.01          & 0.5543    & 0.5575    & 0.5569    & 0.5577   \\ \cline{2-6} 
                           & 0.005                     & 0.5591    & 0.5607    & 0.5610    & 0.5617    \\ \cline{2-6} 
                            & 0.001                    & 0.5621    & 0.5627    & 0.5623    & 0.5639    \\ \cline{2-6}
                           & 0.0005                   & 0.5649    & 0.5661    & 0.5665    & 0.5669    \\ \cline{2-6} 
                           & 0.0001                   & 0.5674    & 0.5675    & 0.5672    & 0.5668    \\ \hline%%%%%%%%%
\multirow{5}{*}{Nadam}     & 0.01       & 0.5672     & 0.5668   & 0.5668   &  0.5658     \\ \cline{2-6} 
			 & 0.005         & 0.5665    & 0.5657    & 0.5644   & 0.5623      \\ \cline{2-6} 
                           & 0.001                    & 0.5621    & 0.5605    & 0.5603   & 0.5583      \\ \cline{2-6} 
                           & 0.0005                  & 0.5576    & 0.5569    & 0.5560   & 0.5540      \\ \cline{2-6} 
                           & 0.0001                  & 0.5526    & 0.5514    & 0.5499   & 0.5485      \\ \hline%%%%%%%
\multirow{5}{*}{Adagrad}   & 0.01        & 0.5486    & 0.5487    & 0.5490    & 0.5490    \\ \cline{2-6} 
			& 0.005           & 0.5492   & 0.5494    & 0.5496    & 0.5497     \\ \cline{2-6} 
                           & 0.001                     & 0.5502   & 0.5503    & 0.5504    & 0.5506    \\ \cline{2-6} 
                           & 0.0005                   & 0.5511   & 0.5516    & 0.5520    & 0.5519    \\ \cline{2-6} 
                           & 0.0001                   & 0.5522   & 0.5524    & 0.5528    & 0.5528     \\ \hline
\end{tabular}
\end{center}
\end{table}
%\vspace{-2 in}
%%%%%%%%%%%%%%%%%%%%%%%%%%%%%%%%%%%%%%%%%%%%%%%%%%%%%%%%%%%%%%

%\newpage

%Hyperparameter tuning for 200 neurons single layer LSTM on the Partial Corr RSFC.

%%%%%%%%%%%%%%%%%%%%%%%%%%%%%%%%%%%%%%%%%%%%%%%%%%%%%%%%
%\textcolor{red}{Table number 1}
\begin{table}[h!]
\begin{center}
	\caption{
	{\bf Hyperparameter tuning for 200 neurons single layer LSTM on the Partial Corr RSFC.}}
	\label{Table:Hyperparameter tuning for 200 neurons single layer LSTM on the Partial Corr RSFC}
		%\resizebox{13cm}{!}{
\begin{tabular}{|c|c|c|c|c|c|}
\hline
\multirow{2}{*}{Optimizer} & \multirow{2}{*}{Learning rate} & \multicolumn{4}{c|}{Batch size} \\ \cline{3-6} 
                           &                                & 4         & 8        & 16    & 32   \\ \hline
\multirow{5}{*}{Adam}      & 0.01          & 0.5650    & 0.5643    & 0.5626    & 0.5602   \\ \cline{2-6} 
                           & 0.005                     & 0.5604    & 0.5589    & 0.5590    & 0.5590    \\ \cline{2-6} 
                            & 0.001                    & 0.5697    & 0.5608    & 0.5614    & 0.5619    \\ \cline{2-6}
                           & 0.0005                   & 0.5636    & 0.5637    & 0.5641    & 0.5639    \\ \cline{2-6} 
                           & 0.0001                   & 0.5646    & 0.5646    & 0.5644    & 0.5638    \\ \hline%%%%%%%%%
\multirow{5}{*}{Nadam}     & 0.01       & 0.5641     & 0.5645   & 0.5654   &  0.5642     \\ \cline{2-6} 
			 & 0.005         & 0.5642    & 0.5642    & 0.5637   & 0.5620      \\ \cline{2-6} 
                           & 0.001                    & 0.5606    & 0.5599    & 0.5581   & 0.5562      \\ \cline{2-6} 
                           & 0.0005                  & 0.5551    & 0.5533    & 0.5521   & 0.5507      \\ \cline{2-6} 
                           & 0.0001                  & 0.5493    & 0.5481    & 0.5469   & 0.5455      \\ \hline%%%%%%%
\multirow{5}{*}{Adagrad}   & 0.01        & 0.5456    & 0.5461    & 0.5463    & 0.5466    \\ \cline{2-6} 
			& 0.005           & 0.5467   & 0.5471    & 0.5474    & 0.5475     \\ \cline{2-6} 
                           & 0.001                     & 0.5478   & 0.5484    & 0.5489    & 0.5490    \\ \cline{2-6} 
                           & 0.0005                   & 0.5495   & 0.5501    & 0.5505    & 0.5508    \\ \cline{2-6} 
                           & 0.0001                   & 0.5515   & 0.5518    & 0.5520    & 0.5519     \\ \hline
\end{tabular}
\end{center}
\end{table}
%\vspace{-2 in}
%%%%%%%%%%%%%%%%%%%%%%%%%%%%%%%%%%%%%%%%%%%%%%%%%%%%%%%%%%%%%%%
%
%\newpage
% 
%%Hyperparameter tuning for 250 neurons single layer LSTM on the Partial Corr RSFC.
%
%%%%%%%%%%%%%%%%%%%%%%%%%%%%%%%%%%%%%%%%%%%%%%%%%%%%%%%%
%\textcolor{red}{Table number 1}
\begin{table}[h!]
\begin{center}
	\caption{
	{\bf Hyperparameter tuning for 250 neurons single layer LSTM on the Partial Corr RSFC.}}
	\label{Table:Hyperparameter tuning for 250 neurons single layer LSTM on the Partial Corr RSFC}
		%\resizebox{13cm}{!}{
\begin{tabular}{|c|c|c|c|c|c|}
\hline
\multirow{2}{*}{Optimizer} & \multirow{2}{*}{Learning rate} & \multicolumn{4}{c|}{Batch size} \\ \cline{3-6} 
                           &                                & 4         & 8        & 16    & 32   \\ \hline
\multirow{5}{*}{Adam}      & 0.01          & 0.5457    & 0.5493    & 0.5524    & 0.5543   \\ \cline{2-6} 
                           & 0.005                     & 0.5536    & 0.5551    & 0.5569    & 0.5565    \\ \cline{2-6} 
                            & 0.001                    & 0.5574    & 0.5589    & 0.5593    & 0.5589    \\ \cline{2-6}
                           & 0.0005                   & 0.5600    & 0.5608    & 0.5604    & 0.5605    \\ \cline{2-6} 
                           & 0.0001                   & 0.5619    & 0.5630    & 0.5626    & 0.5627    \\ \hline%%%%%%%%%
\multirow{5}{*}{Nadam}     & 0.01       & 0.5635     & 0.5646   & 0.5646   &  0.5644     \\ \cline{2-6} 
			 & 0.005         & 0.5643    & 0.5643    & 0.5642   & 0.5627      \\ \cline{2-6} 
                           & 0.001                    & 0.5618    & 0.5605    & 0.5586   & 0.5563      \\ \cline{2-6} 
                           & 0.0005                  & 0.5551    & 0.5535    & 0.5519   & 0.5510      \\ \cline{2-6} 
                           & 0.0001                  & 0.5494    & 0.5481    & 0.5468   & 0.5454      \\ \hline%%%%%%%
\multirow{5}{*}{Adagrad}   & 0.01        & 0.5459    & 0.5462    & 0.5463    & 0.5467    \\ \cline{2-6} 
			& 0.005           & 0.5469   & 0.5470    & 0.5473    & 0.5475     \\ \cline{2-6} 
                           & 0.001                     & 0.5477   & 0.5480    & 0.5484    & 0.5485    \\ \cline{2-6} 
                           & 0.0005                   & 0.5489   & 0.5492    & 0.5493    & 0.5495    \\ \cline{2-6} 
                           & 0.0001                   & 0.5500   & 0.5505    & 0.5509    & 0.5510     \\ \hline
\end{tabular}
\end{center}
\end{table}

%%%%%%%%%%%%%%%%%%%%%%%%%%%%%%%%%%%%%%%%%%%%%%%%%%%%%%%%%%%%%%
\clearpage
\subsection*{Hyperparameter tuning for single layer LSTM on the Spearman's RSFC}
\label{Hyperparameter tuning for single layer LSTM on the Spearman's RSFC}

%Hyperparameter tuning for 10 neurons single layer LSTM on the Spearmans RSFC.

%%%%%%%%%%%%%%%%%%%%%%%%%%%%%%%%%%%%%%%%%%%%%%%%%%%%%%%%
%\textcolor{red}{Table number 1}
\begin{table}[h!]
\begin{center}
	\caption{
	{\bf Hyperparameter tuning for 10 neurons single layer LSTM on the Spearman's RSFC.}}
	\label{Table:Hyperparameter tuning for 10 neurons single layer LSTM on the Spearman's RSFC}
		%\resizebox{13cm}{!}{
\begin{tabular}{|c|c|c|c|c|c|}
\hline
\multirow{2}{*}{Optimizer} & \multirow{2}{*}{Learning rate} & \multicolumn{4}{c|}{Batch size} \\ \cline{3-6} 
                           &                                & 4         & 8        & 16    & 32   \\ \hline
\multirow{5}{*}{Adam}      & 0.01       & 0.6271      & 0.6314    & 0.6424   & 0.6443  \\ \cline{2-6} 
                           & 0.005                  & 0.6427      & 0.6433    & 0.6431   & 0.6441    \\ \cline{2-6} 
                            & 0.001                 & 0.6445      & 0.6442    & 0.6444   & 0.6424    \\ \cline{2-6}
                           & 0.0005                & 0.6426      & 0.6423    & 0.6420   & 0.6422    \\ \cline{2-6} 
                           & 0.0001                & 0.6416      & 0.6406    & 0.6398   & 0.6381   \\ \hline%%%%%%%%%
\multirow{5}{*}{Nadam}     & 0.01      & 0.6372    & 0.6373     & 0.6376   &  0.6380     \\ \cline{2-6} 
			 & 0.005         & 0.6372    & 0.6377     & 0.6374   & 0.6365    \\ \cline{2-6} 
                           & 0.001                  & 0.6353     & 0.6334     & 0.6309   & 0.6284      \\ \cline{2-6} 
                           & 0.0005                & 0.6256     & 0.6231     & 0.6200   & 0.6171    \\ \cline{2-6} 
                           & 0.0001                & 0.6139     & 0.6104     & 0.6072    & 0.6048    \\ \hline%%%%%%%
\multirow{5}{*}{Adagrad}   & 0.01       & 0.6056      & 0.6066     & 0.6075  & 0.6085  \\ \cline{2-6} 
			& 0.005           & 0.6093      & 0.6100     & 0.6110   & 0.6118   \\ \cline{2-6} 
                           & 0.001                   & 0.6125      & 0.6128     & 0.6134    & 0.6141   \\ \cline{2-6} 
                           & 0.0005                 & 0.6150      & 0.6156     & 0.6159    & 0.6162   \\ \cline{2-6} 
                           & 0.0001                 & 0.6163      & 0.6167     & 0.6171   & 0.6172   \\ \hline
\end{tabular}
\end{center}
\end{table}

%%%%%%%%%%%%%%%%%%%%%%%%%%%%%%%%%%%%%%%%%%%%%%%%%%%%%%%%%%%%%

%Hyperparameter tuning for 30 neurons single layer LSTM on the Spearmans RSFC.

%%%%%%%%%%%%%%%%%%%%%%%%%%%%%%%%%%%%%%%%%%%%%%%%%%%%%%%%
%\textcolor{red}{Table number 1}
\begin{table}[h!]
\begin{center}
	\caption{
	{\bf Hyperparameter tuning for 30 neurons single layer LSTM on the Spearman's RSFC.}}
	\label{Table:Hyperparameter tuning for 30 neurons single layer LSTM on the Spearman's RSFC}
		%\resizebox{13cm}{!}{
\begin{tabular}{|c|c|c|c|c|c|}
\hline
\multirow{2}{*}{Optimizer} & \multirow{2}{*}{Learning rate} & \multicolumn{4}{c|}{Batch size} \\ \cline{3-6} 
                           &                                & 4         & 8        & 16    & 32   \\ \hline
\multirow{5}{*}{Adam}      & 0.01       & 0.6286      & 0.6428    & 0.6505   & 0.6504  \\ \cline{2-6} 
                           & 0.005                  & 0.6523      & 0.6529    & 0.6527   & 0.6522    \\ \cline{2-6} 
                            & 0.001                 & 0.6524      & 0.6529    & 0.6528   & 0.6532    \\ \cline{2-6}
                           & 0.0005                & 0.6530      & 0.6539    & 0.6536   & 0.6538    \\ \cline{2-6} 
                           & 0.0001                & 0.6527      & 0.6517    & 0.6512   & 0.6508   \\ \hline%%%%%%%%%
\multirow{5}{*}{Nadam}     & 0.01      & 0.6505    & 0.6502     & 0.6493   &  0.6490     \\ \cline{2-6} 
			 & 0.005         & 0.6486    & 0.6477     & 0.6471   & 0.6461    \\ \cline{2-6} 
                           & 0.001                  & 0.6446     & 0.6432     & 0.6412   & 0.6388     \\ \cline{2-6} 
                           & 0.0005                & 0.6366     & 0.6245     & 0.6317   & 0.6288    \\ \cline{2-6} 
                           & 0.0001                & 0.6265     & 0.6235     & 0.6205    & 0.6175    \\ \hline%%%%%%%
\multirow{5}{*}{Adagrad}   & 0.01       & 0.6180      & 0.6184     & 0.6196   & 0.6201  \\ \cline{2-6} 
			& 0.005           & 0.6207      & 0.6216     & 0.6225   & 0.6234   \\ \cline{2-6} 
                           & 0.001                   & 0.6240      & 0.6243     & 0.6248    & 0.6253   \\ \cline{2-6} 
                           & 0.0005                 & 0.6257      & 0.6261     & 0.6267    & 0.6271   \\ \cline{2-6} 
                           & 0.0001                 & 0.6275      & 0.6273     & 0.6275    & 0.6278   \\ \hline
\end{tabular}
\end{center}
\end{table}

%%%%%%%%%%%%%%%%%%%%%%%%%%%%%%%%%%%%%%%%%%%%%%%%%%%%%%%%%%%%%%
% 
%
%Hyperparameter tuning for 50 neurons single layer LSTM on the Spearmans RSFC.
%
%%%%%%%%%%%%%%%%%%%%%%%%%%%%%%%%%%%%%%%%%%%%%%%%%%%%%%%%
%\textcolor{red}{Table number 1}
\begin{table}[h!]
\begin{center}
	\caption{
	{\bf Hyperparameter tuning for 50 neurons single layer LSTM on the Spearman's RSFC.}}
	\label{Table:Hyperparameter tuning for 50 neurons single layer LSTM on the Spearman's RSFC}
		%\resizebox{13cm}{!}{
\begin{tabular}{|c|c|c|c|c|c|}
\hline
\multirow{2}{*}{Optimizer} & \multirow{2}{*}{Learning rate} & \multicolumn{4}{c|}{Batch size} \\ \cline{3-6} 
                           &                                & 4         & 8        & 16    & 32   \\ \hline
\multirow{5}{*}{Adam}      & 0.01       & 0.6493      & 0.6532    & 0.6626   & 0.6636  \\ \cline{2-6} 
                           & 0.005                  & 0.6641      & 0.6634    & 0.6639   & 0.6635    \\ \cline{2-6} 
                            & 0.001                 & 0.6635      & 0.6629    & 0.6619   & 0.6603    \\ \cline{2-6}
                           & 0.0005                & 0.6597      & 0.6586    & 0.6582   & 0.6571    \\ \cline{2-6} 
                           & 0.0001                & 0.6565      & 0.6562    & 0.6552   & 0.6545   \\ \hline%%%%%%%%%
\multirow{5}{*}{Nadam}     & 0.01      & 0.6540    & 0.6534     & 0.6425   &  0.6516     \\ \cline{2-6} 
			 & 0.005         & 0.6509    & 0.6502     & 0.6497   & 0.6489    \\ \cline{2-6} 
                           & 0.001                  & 0.6482     & 0.6475     & 0.6456   & 0.6435     \\ \cline{2-6} 
                           & 0.0005                & 0.6414     & 0.6391     & 0.6368   & 0.6337    \\ \cline{2-6} 
                           & 0.0001                & 0.6307     & 0.6270     & 0.6242    & 0.6211    \\ \hline%%%%%%%
\multirow{5}{*}{Adagrad}   & 0.01       & 0.6217      & 0.6225     & 0.6234   & 0.6245  \\ \cline{2-6} 
			& 0.005           & 0.6256      & 0.6264     & 0.6270   & 0.6275   \\ \cline{2-6} 
                           & 0.001                   & 0.6283      & 0.6290     & 0.6293    & 0.6296   \\ \cline{2-6} 
                           & 0.0005                 & 0.6303      & 0.6305     & 0.6308    & 0.6312   \\ \cline{2-6} 
                           & 0.0001                 & 0.6315      & 0.6319     & 0.6324    & 0.6326   \\ \hline
\end{tabular}
\end{center}
\end{table}

%%%%%%%%%%%%%%%%%%%%%%%%%%%%%%%%%%%%%%%%%%%%%%%%%%%%%%%%%%%%%%

%Hyperparameter tuning for 100 neurons single layer LSTM on the Spearman's RSFC.

%%%%%%%%%%%%%%%%%%%%%%%%%%%%%%%%%%%%%%%%%%%%%%%%%%%%%%%%
%\textcolor{red}{Table number 1}
\begin{table}[h!]
\begin{center}
	\caption{
	{\bf Hyperparameter tuning for 100 neurons single layer LSTM on the Spearman's RSFC.}}
	\label{Table:Hyperparameter tuning for 100 neurons single layer LSTM on the Spearman's RSFC}
		%\resizebox{13cm}{!}{
\begin{tabular}{|c|c|c|c|c|c|}
\hline
\multirow{2}{*}{Optimizer} & \multirow{2}{*}{Learning rate} & \multicolumn{4}{c|}{Batch size} \\ \cline{3-6} 
                           &                                & 4         & 8        & 16    & 32   \\ \hline
\multirow{5}{*}{Adam}      & 0.01       & 0.6621      & 0.6536    & 0.6505   & 0.6532  \\ \cline{2-6} 
                           & 0.005                  & 0.6524      & 0.6557    & 0.6578   & 0.6571    \\ \cline{2-6} 
                            & 0.001                 & 0.6585      & 0.6591    & 0.6598   & 0.6601    \\ \cline{2-6}
                           & 0.0005                & 0.6605      & 0.6611    & 0.6620   & 0.6615    \\ \cline{2-6} 
                           & 0.0001                & 0.6601      & 0.6604    & 0.6605   & 0.6592   \\ \hline%%%%%%%%%
\multirow{5}{*}{Nadam}     & 0.01      & 0.6586    & 0.6574     & 0.6563   &  0.6551    \\ \cline{2-6} 
			 & 0.005         & 0.6543    & 0.6535     & 0.6524   & 0.6519    \\ \cline{2-6} 
                           & 0.001                  & 0.6510     & 0.6499     & 0.6481   & 0.6468    \\ \cline{2-6} 
                           & 0.0005                & 0.6454     & 0.6446     & 0.6424   & 0.6404    \\ \cline{2-6} 
                           & 0.0001                & 0.6377     & 0.6346     & 0.6327    & 0.6296    \\ \hline%%%%%%%
\multirow{5}{*}{Adagrad}   & 0.01       & 0.6300      & 0.6306     & 0.6312   & 0.6318  \\ \cline{2-6} 
			& 0.005           & 0.6326      & 0.6331     & 0.6336   & 0.6341   \\ \cline{2-6} 
                           & 0.001                   & 0.6350      & 0.6353     & 0.6360    & 0.6365   \\ \cline{2-6} 
                           & 0.0005                 & 0.6369      & 0.6372     & 0.6376    & 0.6379   \\ \cline{2-6} 
                           & 0.0001                 & 0.6383      & 0.6386     & 0.6387    & 0.6388   \\ \hline
\end{tabular}
\end{center}
\end{table}

%%%%%%%%%%%%%%%%%%%%%%%%%%%%%%%%%%%%%%%%%%%%%%%%%%%%%%%%%%%%%%
%Hyperparameter tuning for 150 neurons single layer LSTM on the Spearman's RSFC.

%%%%%%%%%%%%%%%%%%%%%%%%%%%%%%%%%%%%%%%%%%%%%%%%%%%%%%%%
%\textcolor{red}{Table number 1}
\begin{table}[h!]
\begin{center}
	\caption{
	{\bf Hyperparameter tuning for 150 neurons single layer LSTM on the Spearman's RSFC.}}
	\label{Table:Hyperparameter tuning for 150 neurons single layer LSTM on the Spearman's RSFC}
		%\resizebox{13cm}{!}{
\begin{tabular}{|c|c|c|c|c|c|}
\hline
\multirow{2}{*}{Optimizer} & \multirow{2}{*}{Learning rate} & \multicolumn{4}{c|}{Batch size} \\ \cline{3-6} 
                           &                                & 4         & 8        & 16    & 32   \\ \hline
\multirow{5}{*}{Adam}      & 0.01       & 0.6336      & 0.6343    & 0.6374   & 0.6359  \\ \cline{2-6} 
                           & 0.005                  & 0.6364      & 0.6363    & 0.6361   & 0.6361    \\ \cline{2-6} 
                            & 0.001                 & 0.6379      & 0.6406    & 0.6417   & 0.6419    \\ \cline{2-6}
                           & 0.0005                & 0.6427      & 0.6433    & 0.6437   & 0.6442    \\ \cline{2-6} 
                           & 0.0001                & 0.6447      & 0.6453    & 0.6454   & 0.6460   \\ \hline%%%%%%%%%
\multirow{5}{*}{Nadam}     & 0.01      & 0.6462    & 0.6460     & 0.6465   &  0.6468    \\ \cline{2-6} 
			 & 0.005         & 0.6473    & 0.6475     & 0.6472   & 0.6470    \\ \cline{2-6} 
                           & 0.001                  & 0.6471     & 0.6470     & 0.6463   & 0.6452    \\ \cline{2-6} 
                           & 0.0005                & 0.6446     & 0.6434     & 0.6416   & 0.6391    \\ \cline{2-6} 
                           & 0.0001                & 0.6365     & 0.6341     & 0.6315    & 0.6282    \\ \hline%%%%%%%
\multirow{5}{*}{Adagrad}   & 0.01       & 0.6282      & 0.6285     & 0.6288   & 0.6290  \\ \cline{2-6} 
			& 0.005           & 0.6293      & 0.6294     & 0.6298   & 0.6302   \\ \cline{2-6} 
                           & 0.001                   & 0.6309      & 0.6312     & 0.6314    & 0.6318   \\ \cline{2-6} 
                           & 0.0005                 & 0.6322      & 0.6326     & 0.6329    & 0.6333   \\ \cline{2-6} 
                           & 0.0001                 & 0.6337      & 0.6340     & 0.6341    & 0.6345   \\ \hline
\end{tabular}
\end{center}
\end{table}

%%%%%%%%%%%%%%%%%%%%%%%%%%%%%%%%%%%%%%%%%%%%%%%%%%%%%%%%%%%%%%
%Hyperparameter tuning for 200 neurons single layer LSTM on the Spearman's RSFC.

%%%%%%%%%%%%%%%%%%%%%%%%%%%%%%%%%%%%%%%%%%%%%%%%%%%%%%%%
%\textcolor{red}{Table number 1}
\begin{table}[h!]
\begin{center}
	\caption{
	{\bf Hyperparameter tuning for 200 neurons single layer LSTM on the Spearman's RSFC.}}
	\label{Table:Hyperparameter tuning for 200 neurons single layer LSTM on the Spearman's RSFC}
		%\resizebox{13cm}{!}{
\begin{tabular}{|c|c|c|c|c|c|}
\hline
\multirow{2}{*}{Optimizer} & \multirow{2}{*}{Learning rate} & \multicolumn{4}{c|}{Batch size} \\ \cline{3-6} 
                           &                                & 4         & 8        & 16    & 32   \\ \hline
\multirow{5}{*}{Adam}      & 0.01       & 0.6279      & 0.6354    & 0.6345   & 0.6348  \\ \cline{2-6} 
                           & 0.005                  & 0.6347      & 0.6361    & 0.6360   & 0.6382    \\ \cline{2-6} 
                            & 0.001                 & 0.6397      & 0.6402    & 0.6416   & 0.6420    \\ \cline{2-6}
                           & 0.0005                & 0.6432      & 0.6434    & 0.6443   & 0.6451    \\ \cline{2-6} 
                           & 0.0001                & 0.6462      & 0.6466    & 0.6469   & 0.6470   \\ \hline%%%%%%%%%
\multirow{5}{*}{Nadam}     & 0.01      & 0.6471    & 0.6478     & 0.6484   &  0.6484    \\ \cline{2-6} 
			 & 0.005         & 0.6481    & 0.6485     & 0.6483   & 0.6486    \\ \cline{2-6} 
                           & 0.001                  & 0.6491     & 0.6493     & 0.6490   & 0.6477    \\ \cline{2-6} 
                           & 0.0005                & 0.6478     & 0.6472     & 0.6459   & 0.6439    \\ \cline{2-6} 
                           & 0.0001                & 0.6410     & 0.6382     & 0.6350    & 0.6323    \\ \hline%%%%%%%
\multirow{5}{*}{Adagrad}   & 0.01       & 0.6324      & 0.6323     & 0.6330   & 0.6331  \\ \cline{2-6} 
			& 0.005           & 0.6330      & 0.6331     & 0.6335   & 0.6340  \\ \cline{2-6} 
                           & 0.001                   & 0.6342      & 0.6346     & 0.6349    & 0.6351   \\ \cline{2-6} 
                           & 0.0005                 & 0.6353      & 0.6357     & 0.6361    & 0.6363   \\ \cline{2-6} 
                           & 0.0001                 & 0.6367      & 0.6371     & 0.6372    & 0.6375   \\ \hline
\end{tabular}
\end{center}
\end{table}

%%%%%%%%%%%%%%%%%%%%%%%%%%%%%%%%%%%%%%%%%%%%%%%%%%%%%%%%%%%%%%
%Hyperparameter tuning for 250 neurons single layer LSTM on the Spearman's RSFC.

%%%%%%%%%%%%%%%%%%%%%%%%%%%%%%%%%%%%%%%%%%%%%%%%%%%%%%%%
%\textcolor{red}{Table number 1}
\begin{table}[h!]
\begin{center}
	\caption{
	{\bf Hyperparameter tuning for 250 neurons single layer LSTM on the Spearman's RSFC.}}
	\label{Table:Hyperparameter tuning for 250 neurons single layer LSTM on the Spearman's RSFC}
		%\resizebox{13cm}{!}{
\begin{tabular}{|c|c|c|c|c|c|}
\hline
\multirow{2}{*}{Optimizer} & \multirow{2}{*}{Learning rate} & \multicolumn{4}{c|}{Batch size} \\ \cline{3-6} 
                           &                                & 4         & 8        & 16    & 32   \\ \hline
\multirow{5}{*}{Adam}      & 0.01       & 0.6400      & 0.6379    & 0.6379   & 0.6391  \\ \cline{2-6} 
                           & 0.005                  & 0.6377      & 0.6362    & 0.6376   & 0.6377    \\ \cline{2-6} 
                            & 0.001                 & 0.6390      & 0.6411    & 0.6425   & 0.6425    \\ \cline{2-6}
                           & 0.0005                & 0.6436      & 0.6447    & 0.6446   & 0.6450    \\ \cline{2-6} 
                           & 0.0001                & 0.6453      & 0.6454    & 0.6456   & 0.6459   \\ \hline%%%%%%%%%
\multirow{5}{*}{Nadam}     & 0.01      & 0.6467    & 0.6468     & 0.6472   &  0.6479    \\ \cline{2-6} 
			 & 0.005         & 0.6478    & 0.6485     & 0.6479   & 0.6480    \\ \cline{2-6} 
                           & 0.001                  & 0.6484     & 0.6489     & 0.6487   & 0.6476    \\ \cline{2-6} 
                           & 0.0005                & 0.6476     & 0.6475     & 0.6462   & 0.6443    \\ \cline{2-6} 
                           & 0.0001                & 0.6417     & 0.6391     & 0.6366    & 0.6336    \\ \hline%%%%%%%
\multirow{5}{*}{Adagrad}   & 0.01       & 0.6335      & 0.6334     & 0.6337   & 0.6337  \\ \cline{2-6} 
			& 0.005           & 0.6339      & 0.6341     & 0.6341   & 0.6345  \\ \cline{2-6} 
                           & 0.001                   & 0.6348      & 0.6350     & 0.6355    & 0.6359   \\ \cline{2-6} 
                           & 0.0005                 & 0.6364      & 0.6367     & 0.6371    & 0.6375   \\ \cline{2-6} 
                           & 0.0001                 & 0.6376      & 0.6378     & 0.6381    & 0.6384   \\ \hline
\end{tabular}
\end{center}
\end{table}

%%%%%%%%%%%%%%%%%%%%%%%%%%%%%%%%%%%%%%%%%%%%%%%%%%%%%%%%%%%%%%
\clearpage
\subsection{\large\bf  Hyperparameter tuning for single layer GRU with various RSFCs}
\label{Hyperparameter tuning for single layer GRU with various RSFCs}

\subsection*{ Hyperparameter tuning for single layer GRU on the Pearson’s RSFC}
%{\large\bf Hyperparameter tuning for single layer GRU on the Pearson’s RSFC }
%%%%%%%%%%%%%%%%%%%%%%%%%%%%%%%%%%%%%%%%%%%%%%%%%%%%%%%%%%%%%%

%%%%%%%%%%%%%%%%%%%%%%%%%%%%%%%%%%%%%%%%%%%%%%%%%%%%%%%%
%\textcolor{red}{Table number 1}
\begin{table}[h!]
\begin{center}
	\caption{
	{\bf Hyperparameter tuning for 10 neurons single layer GRU with Pearson’s RSFC.}}
	\label{Table:Hyperparameter tuning for 10 neurons single layer GRU with Pearson’s RSFC}
		%\resizebox{13cm}{!}{
\begin{tabular}{|c|c|c|c|c|c|}
\hline
\multirow{2}{*}{Optimizer} & \multirow{2}{*}{Learning rate} & \multicolumn{4}{c|}{Batch size} \\ \cline{3-6} 
                           &                                & 4         & 8        & 16    & 32   \\ \hline
\multirow{4}{*}{Adam}      & 0.01                            & 0.6085     & 0.6160    & 0.6169 & 0.6155 \\ \cline{2-6} 
                           & 0.001                           & 0.6161       & 0.6145 &  0.6144 &  0.6134  \\ \cline{2-6} 
                           & 0.0005                          & 0.6129      & 0.6134       & 0.6145  & 0.6152    \\ \cline{2-6} 
                           & 0.0001                          & 0.6163       & \bf{0.6171}      & 0.6165 & 0.6152     \\ \hline
\multirow{4}{*}{Nadam}     & 0.01                            & 0.6150        & 0.6158       & 0.6156 & 0.6167       \\ \cline{2-6} 
                           & 0.001                           & 0.6141      & 0.6111       & 0.6092    & 0.6058   \\ \cline{2-6} 
                           & 0.0005                           &0.6017         & 0.5994       & 0.5966 &   0.5934   \\ \cline{2-6} 
                           & 0.0001                          & 0.5986      & 0.5879       & 0.5848 & 0.5823      \\ \hline
\multirow{4}{*}{Adagrad}   & 0.01                            &0.5829        & 0.5836      & 0.5843 & 0.5854     \\ \cline{2-6} 
                           & 0.001                           & 0.5862      & 0.5870     & 0.5878 & 0.5886     \\ \cline{2-6} 
                           & 0.0005                          & 0.5891     & 0.5899       & 0.5905 & 0.5913     \\ \cline{2-6} 
                           & 0.0001                          & 0.5921        & 0.5923       & 0.5926 &   0.5921   \\ \hline
\end{tabular}
\end{center}
\end{table}

%%%%%%%%%%%%%%%%%%%%%%%%%%%%%%%%%%%%%%%%%%%%%%%%%%%%%%%%%%%%%%

%%%%%%%%%%%%%%%%%%%%%%%%%%%%%%%%%%%%%%%%%%%%%%%%%%%%%%%%
%\textcolor{red}{Table number 2}
\begin{table}[h!]
\begin{center}
	\caption{
	{\bf Hyperparameter tuning for 30 neurons single layer GRU with Pearson’s RSFC.}}
	\label{Table:Hyperparameter tuning for 30 neurons single layer GRU with Pearson’s RSFC}
		%\resizebox{13cm}{!}{
\begin{tabular}{|c|c|c|c|c|c|}
\hline
\multirow{2}{*}{Optimizer} & \multirow{2}{*}{Learning rate} & \multicolumn{4}{c|}{Batch size} \\ \cline{3-6} 
                           &                                & 4         & 8        & 16    & 32   \\ \hline
\multirow{4}{*}{Adam}      & 0.01                            & 0.6029 & 0. .6132    & 0.6160 & 0.6170 \\ \cline{2-6} 
                           & 0.001                           & 0. 6186       & 0. 6188 &  0. 6187 &  0. 6199  \\ \cline{2-6} 
                           & 0.0005                          & 0. 6209      & 0. 6204       & 0. 6214  & 0. 6213    \\ \cline{2-6} 
                           & 0.0001                          & 0. 6217       & \bf{0. 6215}      & 0. 6210 & 0. 6202     \\ \hline
\multirow{4}{*}{Nadam}     & 0.01                            & 0. 6205        & 0. 6206       & 0. 6202 & 0. 6191       \\ \cline{2-6} 
                           & 0.001                           & 0. 6182      & 0. 6167       & 0. 6152    & 0. 6125   \\ \cline{2-6} 
                           & 0.0005                           &0. 6104         & 0. 6081       & 0. 6059 &   0. 6028   \\ \cline{2-6} 
                           & 0.0001                          & 0. 5994      & 0. 5965       & 0. 5939 & 0. 5912      \\ \hline
\multirow{4}{*}{Adagrad}   & 0.01                            &0. 5916        & 0. 5923      & 0. 5930 & 0. 5936     \\ \cline{2-6} 
                           & 0.001                           & 0. 5945      & 0. 5951     & 0. 5957 & 0. 5964     \\ \cline{2-6} 
                           & 0.0005                          & 0. 5967     & 0. 5974       & 0. 5980 & 0. 5985     \\ \cline{2-6} 
                           & 0.0001                          & 0. 5993        & 0. 5999       & 0. 6003 &   0. 6004   \\ \hline
\end{tabular}
\end{center}
\end{table}

%%%%%%%%%%%%%%%%%%%%%%%%%%%%%%%%%%%%%%%%%%%%%%%%%%%%%%%%%%%%%%

%%%%%%%%%%%%%%%%%%%%%%%%%%%%%%%%%%%%%%%%%%%%%%%%%%%%%%%%
%\textcolor{red}{Table number 3}
\begin{table}[h!]
\begin{center}
	\caption{
	{\bf Hyperparameter tuning for 50 neurons single layer GRU with Pearson’s RSFC.}}
	\label{Table:Hyperparameter tuning for 50 neurons single layer GRU with Pearson’s RSFC}
		%\resizebox{13cm}{!}{
\begin{tabular}{|c|c|c|c|c|c|}
\hline
\multirow{2}{*}{Optimizer} & \multirow{2}{*}{Learning rate} & \multicolumn{4}{c|}{Batch size} \\ \cline{3-6} 
                           &                                & 4         & 8        & 16    & 32   \\ \hline
\multirow{4}{*}{Adam}      & 0.01                            & 0. 6107 & 0. . 6125    & 0. 6174 & 0. 6152 \\ \cline{2-6} 
                           & 0.001                           & 0. 6187       & 0. 62 &  0. 62 &  0. 6208  \\ \cline{2-6} 
                           & 0.0005                          & 0. 6216      & 0. 6221       & 0. 6232  & 0. 6227    \\ \cline{2-6} 
                           & 0.0001                          & 0. 6229       & \bf{0. 6220}      & 0. 6222 & 0. 6233     \\ \hline
\multirow{4}{*}{Nadam}     & 0.01                            & 0. 6234        & 0. 6230       & 0. 6229 & 0. 6230       \\ \cline{2-6} 
                           & 0.001                           & 0. 6224      & 0. 6209       & 0. 6192    & 0. 6174   \\ \cline{2-6} 
                           & 0.0005                           &0. 616         & 0. 6143       & 0. 6120 &   0. 6093   \\ \cline{2-6} 
                           & 0.0001                          & 0. 6064      & 0. 6030       & 0. 5999 & 0. 5970      \\ \hline
\multirow{4}{*}{Adagrad}   & 0.01                            &0. 5977        & 0. 5984      & 0. 5988 & 0. 5992     \\ \cline{2-6} 
                           & 0.001                           & 0. 5999      & 0. 6006     & 0. 6013 & 0. 6018     \\ \cline{2-6} 
                           & 0.0005                          & 0. 6023     & 0. 6028       & 0. 6032 & 0. 6038     \\ \cline{2-6} 
                           & 0.0001                          & 0. 6044        & 0. 6051       & 0. 6056 &   0. 6058   \\ \hline
\end{tabular}
\end{center}
\end{table}

%%%%%%%%%%%%%%%%%%%%%%%%%%%%%%%%%%%%%%%%%%%%%%%%%%%%%%%%
%\textcolor{red}{Table number 4}
\begin{table}[h!]
\begin{center}
	\caption{
	{\bf Hyperparameter tuning for 100 neurons single layer GRU with Pearson’s RSFC.}}
	\label{Table:Hyperparameter tuning for 100 neurons single layer GRU with Pearson’s RSFC}
		%\resizebox{13cm}{!}{
\begin{tabular}{|c|c|c|c|c|c|}
\hline
\multirow{2}{*}{Optimizer} & \multirow{2}{*}{Learning rate} & \multicolumn{4}{c|}{Batch size} \\ \cline{3-6} 
                           &                                & 4         & 8        & 16    & 32   \\ \hline
\multirow{4}{*}{Adam}      & 0.01                            & 0. 6157 & 0. . 6154    & 0. 6114 & 0. 6093 \\ \cline{2-6} 
                           & 0.001                           & 0. 6129       & 0. 6146 &  0. 6155 &  0. 6159  \\ \cline{2-6} 
                           & 0.0005                          & 0. 6175      & 0. 6188       & 0. 6190  & 0. 6189    \\ \cline{2-6} 
                           & 0.0001                          & 0. 6195       & \bf{0. 6198}      & 0. 6196 & 0. 6197     \\ \hline
\multirow{4}{*}{Nadam}     & 0.01                            & 0. 6204        & 0. 6208       & 0. 6213 & 0. 6211       \\ \cline{2-6} 
                           & 0.001                           & 0. 6207      & 0. 6200       & 0. 6194    & 0. 6183   \\ \cline{2-6} 
                           & 0.0005                           &0. 6171         & 0. 6148       & 0. 6132 &   0. 6110   \\ \cline{2-6} 
                           & 0.0001                          & 0. 6086      & 0. 6055       & 0. 6023 & 0. 5997      \\ \hline
\multirow{4}{*}{Adagrad}   & 0.01                            &0. 6000        & 0. 6009      & 0. 6013 & 0. 6018     \\ \cline{2-6} 
                           & 0.001                           & 0. 6022      & 0. 6025     & 0. 6030 & 0. 6036     \\ \cline{2-6} 
                           & 0.0005                          & 0. 6040     & 0. 6046       & 0. 6050 & 0. 6054     \\ \cline{2-6} 
                           & 0.0001                          & 0. 6060        & 0. 6064       & 0. 6064 &   0. 6066   \\ \hline
\end{tabular}
\end{center}
\end{table}

%%%%%%%%%%%%%%%%%%%%%%%%%%%%%%%%%%%%%%%%%%%%%%%%%%%%%%%%%%%%%%

%%%%%%%%%%%%%%%%%%%%%%%%%%%%%%%%%%%%%%%%%%%%%%%%%%%%%%%%
%\textcolor{red}{Table number 5}
\begin{table}[h!]
\begin{center}
	\caption{
	{\bf Hyperparameter tuning for 150 neurons single layer GRU with Pearson’s RSFC.}}
	\label{Table:Hyperparameter tuning for 150 neurons single layer GRU with Pearson’s RSFC}
		%\resizebox{13cm}{!}{
\begin{tabular}{|c|c|c|c|c|c|}
\hline
\multirow{2}{*}{Optimizer} & \multirow{2}{*}{Learning rate} & \multicolumn{4}{c|}{Batch size} \\ \cline{3-6} 
                           &                                & 4         & 8        & 16    & 32   \\ \hline
\multirow{4}{*}{Adam}      & 0.01                            & 0.6193 & 0.6179    & 0.6140 & 0.6104 \\ \cline{2-6} 
                           & 0.001                           & 0.6129       & 0.6157 &  0.6169 &  0.6167  \\ \cline{2-6} 
                           & 0.0005                          & 0.6181      & 0.6181       & 0.6188  & 0.6198    \\ \cline{2-6} 
                           & 0.0001                          & 0.6197       & \bf{0.62}      & 0.6199 & 0.6212     \\ \hline
\multirow{4}{*}{Nadam}     & 0.01                            & 0.6213        & 0.6215       & 0.6213 & 0. 6205        \\ \cline{2-6} 
                           & 0.001                           & 0.6213      & 0.6222       & 0.6216    & 0.6201   \\ \cline{2-6} 
                           & 0.0005                           &0.6194         & 0.6180       & 0.6176 &   0.6157   \\ \cline{2-6} 
                           & 0.0001                          & 0.6132      & 0.6101       & 0.6073 & 0.6035      \\ \hline
\multirow{4}{*}{Adagrad}   & 0.01                            &0.6036        & 0.6038      & 0.6046 & 0. 6049     \\ \cline{2-6} 
                           & 0.001                           & 0.6056      & 0.6064     & 0.6068 & 0.6071     \\ \cline{2-6} 
                           & 0.0005                          & 0.6077     & 0.6081       & 0.6083 & 0.6087     \\ \cline{2-6} 
                           & 0.0001                          & 0.6088        & 0.6091       & 0.6093 &   0.6097   \\ \hline
\end{tabular}
\end{center}
\end{table}

%%%%%%%%%%%%%%%%%%%%%%%%%%%%%%%%%%%%%%%%%%%%%%%%%%%%%%%%
%\textcolor{red}{Table number 6}
\begin{table}[h!]
\begin{center}
	\caption{
	{\bf Hyperparameter tuning for 200 neurons single layer GRU with Pearson’s RSFC.}}
	\label{Table:Hyperparameter tuning for 200 neurons single layer GRU with Pearson’s RSFC}
		%\resizebox{13cm}{!}{
\begin{tabular}{|c|c|c|c|c|c|}
\hline
\multirow{2}{*}{Optimizer} & \multirow{2}{*}{Learning rate} & \multicolumn{4}{c|}{Batch size} \\ \cline{3-6} 
                           &                                & 4         & 8        & 16    & 32   \\ \hline
\multirow{4}{*}{Adam}      & 0.01                            & 0. 6114 & 0. . 6168    & 0. 6186 & 0. 6145 \\ \cline{2-6} 
                           & 0.001                           & 0. 6191       & 0. 6202 &  0. 6208 &  0. 6221  \\ \cline{2-6} 
                           & 0.0005                          & 0. 6231      & 0. 6246       & 0. 6242  & 0. 6242    \\ \cline{2-6} 
                           & 0.0001                          & 0. 6242       & \bf{0. 6236}      & 0. 6236 & 0. 6229     \\ \hline
\multirow{4}{*}{Nadam}     & 0.01                            & 0. 6227        & 0. 6227       & 0. 6223 & 0. 6217        \\ \cline{2-6} 
                           & 0.001                           & 0. 6222      & 0. 6224       & 0. 6209    & 0. 6207   \\ \cline{2-6} 
                           & 0.0005                           &0. 6206         & 0. 6194       & 0. 6178 &   0. 6173   \\ \cline{2-6} 
                           & 0.0001                          & 0. 6146 & 0. 6111       & 0. 6081 & 0. 6052      \\ \hline
\multirow{4}{*}{Adagrad}   & 0.01                            &0. 6055        & 0. 6058      & 0. 6060 & 0. 6063     \\ \cline{2-6} 
                           & 0.001                           & 0. 6071      & 0. 6077     & 0. 6081 & 0. 6086     \\ \cline{2-6} 
                           & 0.0005                          & 0. 6092     & 0. 6098       & 0. 6102 & 0. 6106     \\ \cline{2-6} 
                           & 0.0001                          & 0. 6120        & 0. 6111       & 0. 6115 &   0. 6115   \\ \hline
\end{tabular}
\end{center}
\end{table}

%%%%%%%%%%%%%%%%%%%%%%%%%%%%%%%%%%%%%%%%%%%%%%%%%%%%%%%%

%%%%%%%%%%%%%%%%%%%%%%%%%%%%%%%%%%%%%%%%%%%%%%%%%%%%%%%%
%\textcolor{red}{Table number 7}
\begin{table}[h!]
\begin{center}
	\caption{
	{\bf Hyperparameter tuning for 250 neurons single layer GRU with Pearson’s RSFC.}}
	\label{Table:Hyperparameter tuning for 250 neurons single layer GRU with Pearson’s RSFC}
		%\resizebox{13cm}{!}{
\begin{tabular}{|c|c|c|c|c|c|}
\hline
\multirow{2}{*}{Optimizer} & \multirow{2}{*}{Learning rate} & \multicolumn{4}{c|}{Batch size} \\ \cline{3-6} 
                           &                                & 4         & 8        & 16    & 32   \\ \hline
\multirow{4}{*}{Adam}      & 0.01                            & 0. 6193     & 0. 6168    & 0. 6176 & 0. 6171 \\ \cline{2-6} 
                           & 0.001                           & 0. 6191       & 0. 6218 &  0. 6216 &  0. 6221  \\ \cline{2-6} 
                           & 0.0005                          & 0. 6230      & 0. 6236       & 0. 6242  & 0. 6242    \\ \cline{2-6} 
                           & 0.0001                          & 0. 6242       & \bf{0. 6241}      & 0. 6234 & 0. 6230     \\ \hline
\multirow{4}{*}{Nadam}     & 0.01                            & 0. 6235        & 0. 6236       & 0. 6229 & 0. 6235       \\ \cline{2-6} 
                           & 0.001                           & 0. 6233      & 0. 6244       & 0. 6245    & 0. 6233   \\ \cline{2-6} 
                           & 0.0005                           &0. 6240         & 0. 6235       & 0. 6227 &   0. 6204   \\ \cline{2-6} 
                           & 0.0001                          & 0. 6184      & 0. 6157       & 0. 6132 & 0. 6101      \\ \hline
\multirow{4}{*}{Adagrad}   & 0.01                            &0. 6111        & 0. 6116      & 0. 6123 & 0. 6127     \\ \cline{2-6} 
                           & 0.001                           & 0. 6132      & 0. 6138     & 0. 6142 & 0. 6144     \\ \cline{2-6} 
                           & 0.0005                          & 0. 6146     & 0. 6148       & 0. 6152 & 0. 6155     \\ \cline{2-6} 
                           & 0.0001                          & 0. 6156        & 0. 6156       & 0. 6160 &   0. 6160   \\ \hline
\end{tabular}
\end{center}
\end{table}

%%%%%%%%%%%%%%%%%%%%%%%%%%%%%%%%%%%%%%%%%%%%%%%%%%%%%%%%%%%%%%
%%%%%%%%%%%%%%%%%%%%%%%%%%%%%%%%%%%%%%%%%%%%%%%%%%%%%%%%%%%%%%
\clearpage
\subsection*{ Hyperparameter tuning for single layer GRU on the Spearman’s RSFC}

%%%%%%%%%%%%%%%%%%%%%%%%%%%%%%%%%%%%%%%%%%%%%%%%%%%%%%%%%%%%%%

%%%%%%%%%%%%%%%%%%%%%%%%%%%%%%%%%%%%%%%%%%%%%%%%%%%%%%%%
%\textcolor{red}{Table number 1}
\begin{table}[h!]
\begin{center}
	\caption{
	{\bf Hyperparameter tuning for 10 neurons single layer GRU with Spearmans RSFC.}}
	\label{Table:Hyperparameter tuning for 10 neurons single layer GRU with Spearmans RSFC}
		%\resizebox{13cm}{!}{
\begin{tabular}{|c|c|c|c|c|c|}
\hline
\multirow{2}{*}{Optimizer} & \multirow{2}{*}{Learning rate} & \multicolumn{3}{c|}{Batch size} \\ \cline{3-5} 
                           &                                & 4         & 8        & 16    \\ \hline
\multirow{4}{*}{Adam}      & 0.01                            & 0. 6236     & 0. 6225    & 0. 6229 \\ \cline{2-5} 
                           & 0.001                           & 0. 6211       & 0. 6226 &  0. 6230 \\ \cline{2-5} 
                           & 0.0005                          & 0. 6215      & 0. 6220       & 0. 6214  \\ \cline{2-5} 
                           & 0.0001                          & 0. 6212       & \bf{0. 6207}      & 0. 6187 \\ \hline
\multirow{4}{*}{Nadam}     & 0.01                            & 0. 6197        & 0. 6209       & 0. 6206 \\ \cline{2-5} 
                           & 0.001                           & 0. 6174      & 0. 6137       & 0. 6112    \\ \cline{2-5} 
                           & 0.0005                           &0. 6062         & 0. 6021       & 0. 5981 \\ \cline{2-5} 
                           & 0.0001                          & 0. 5947      & 0. 5907       & 0. 5870 \\ \hline
\multirow{4}{*}{Adagrad}   & 0.01                            &0. 5889        & 0. 5902      & 0. 5913 \\ \cline{2-5} 
                           & 0.001                           & 0. 5924      & 0. 5936     & 0.5946 \\ \cline{2-5} 
                           & 0.0005                          & 0. 5953     & 0. 5962       & 0. 5968 \\ \cline{2-5} 
                           & 0.0001                          & 0. 5981        & 0. 5988       & 0. 5990 \\ \hline
\end{tabular}
\end{center}
\end{table}

%%%%%%%%%%%%%%%%%%%%%%%%%%%%%%%%%%%%%%%%%%%%%%%%%%%%%%%%%%%%%%
%%%%%%%%%%%%%%%%%%%%%%%%%%%%%%%%%%%%%%%%%%%%%%%%%%%%%%%%
%\textcolor{red}{Table number 2}
\begin{table}[h!]
\begin{center}
	\caption{
	{\bf Hyperparameter tuning for 30 neurons single layer GRU with Spearmans RSFC.}}
	\label{Table:Hyperparameter tuning for 30 neurons single layer GRU with Spearmans RSFC}
		%\resizebox{13cm}{!}{
\begin{tabular}{|c|c|c|c|c|c|}
\hline
\multirow{2}{*}{Optimizer} & \multirow{2}{*}{Learning rate} & \multicolumn{3}{c|}{Batch size} \\ \cline{3-5} 
                           &                                & 4         & 8        & 16    \\ \hline
\multirow{4}{*}{Adam}      & 0.01                            & 0. 6193     & 0. 6254    & 0. 6271 \\ \cline{2-5} 
                           & 0.001                           & 0. 6254       & 0. 625 &  0. 6254 \\ \cline{2-5} 
                           & 0.0005                          & 0. 6261      & 0. 6253       & 0. 6249  \\ \cline{2-5} 
                           & 0.0001                          & 0. 6272      & \bf{0. 6272}      & 0. 6265 \\ \hline
\multirow{4}{*}{Nadam}     & 0.01                            & 0. 6261        & 0. 6256       & 0. 6259 \\ \cline{2-5} 
                           & 0.001                           & 0. 6240      & 0. 6218       & 0. 6199    \\ \cline{2-5} 
                           & 0.0005                           &0. 6169         & 0. 6133       & 0. 6095 \\ \cline{2-5} 
                           & 0.0001                          & 0. 6056      & 0. 6007       & 0. 5966 \\ \hline
\multirow{4}{*}{Adagrad}   & 0.01                            &0. 5979       & 0. 5983      & 0. 5994 \\ \cline{2-5} 
                           & 0.001                           & 0. 6003      & 0. 6011     & 0. 6017 \\ \cline{2-5} 
                           & 0.0005                          & 0. 6029     & 0. 6038       & 0. 6044 \\ \cline{2-5} 
                           & 0.0001                          & 0. 6051        & 0. 6058       & 0. 6064 \\ \hline
\end{tabular}
\end{center}
\end{table}

%%%%%%%%%%%%%%%%%%%%%%%%%%%%%%%%%%%%%%%%%%%%%%%%%%%%%%%%%%%%%%
%%%%%%%%%%%%%%%%%%%%%%%%%%%%%%%%%%%%%%%%%%%%%%%%%%%%%%%%
%\textcolor{red}{Table number 3}
\begin{table}[h!]
\begin{center}
	\caption{
	{\bf Hyperparameter tuning for 50 neurons single layer GRU with Spearmans RSFC.}}
	\label{Table:Hyperparameter tuning for 50 neurons single layer GRU with Spearmans RSFC}
		%\resizebox{13cm}{!}{
\begin{tabular}{|c|c|c|c|c|c|}
\hline
\multirow{2}{*}{Optimizer} & \multirow{2}{*}{Learning rate} & \multicolumn{3}{c|}{Batch size} \\ \cline{3-5} 
                           &                                & 4         & 8        & 16    \\ \hline
\multirow{4}{*}{Adam}      & 0.01                            & 0. 6164     & 0. 6214    & 0. 6212 \\ \cline{2-5} 
                           & 0.001                           & 0. 6230       & 0. 6219 &  0. 6227 \\ \cline{2-5} 
                           & 0.0005                          & 0. 6224      & 0. 6219       & 0. 6213  \\ \cline{2-5} 
                           & 0.0001                          & 0. 6219      & \bf{0. 6223}      & 0. 6229 \\ \hline
\multirow{4}{*}{Nadam}     & 0.01                            & 0. 6235        & 0. 6244      & 0. 6252 \\ \cline{2-5} 
                           & 0.001                           & 0. 6241      & 0. 6218       & 0. 6194    \\ \cline{2-5} 
                           & 0.0005                           &0. 6161         & 0. 6136       & 0. 6121 \\ \cline{2-5} 
                           & 0.0001                          & 0. 6080      & 0. 6043       & 0. 6009 \\ \hline
\multirow{4}{*}{Adagrad}   & 0.01                            &0. 6019       & 0. 6024     & 0. 6034 \\ \cline{2-5} 
                           & 0.001                           & 0. 6043      & 0. 6050     & 0. 6055 \\ \cline{2-5} 
                           & 0.0005                          & 0. 6059     & 0. 6063       & 0. 6065 \\ \cline{2-5} 
                           & 0.0001                          & 0. 6071        & 0. 6076       & 0. 6084 \\ \hline
\end{tabular}
\end{center}
\end{table}

%%%%%%%%%%%%%%%%%%%%%%%%%%%%%%%%%%%%%%%%%%%%%%%%%%%%%%%%%%%%%%
%%%%%%%%%%%%%%%%%%%%%%%%%%%%%%%%%%%%%%%%%%%%%%%%%%%%%%%%
%\textcolor{red}{Table number 4}
\begin{table}[h!]
\begin{center}
	\caption{
	{\bf Hyperparameter tuning for 100 neurons single layer GRU with Spearmans RSFC.}}
	\label{Table:Hyperparameter tuning for 100 neurons single layer GRU with Spearmans RSFC}
		%\resizebox{13cm}{!}{
\begin{tabular}{|c|c|c|c|c|c|}
\hline
\multirow{2}{*}{Optimizer} & \multirow{2}{*}{Learning rate} & \multicolumn{3}{c|}{Batch size} \\ \cline{3-5} 
                           &                                & 4         & 8        & 16    \\ \hline
\multirow{4}{*}{Adam}      & 0.01                            & 0. 6286     & 0. 6264    & 0. 6295 \\ \cline{2-5} 
                           & 0.001                           & 0. 6266       & 0. 6251 &  0. 6258 \\ \cline{2-5} 
                           & 0.0005                          & 0. 6259      & 0. 6253       & 0. 6248  \\ \cline{2-5} 
                           & 0.0001                          & 0. 6243     & \bf{0. 6232}      & 0. 6240 \\ \hline
\multirow{4}{*}{Nadam}     & 0.01                            & 0. 6242        & 0. 6246     & 0. 625 \\ \cline{2-5} 
                           & 0.001                           & 0. 6236      & 0. 6235       & 0. 6217    \\ \cline{2-5} 
                           & 0.0005                           &0. 6193         & 0. 6171       & 0. 6143 \\ \cline{2-5} 
                           & 0.0001                          & 0. 6096      & 0. 6060       & 0. 6026 \\ \hline
\multirow{4}{*}{Adagrad}   & 0.01                            &0. 6035       & 0. 6041     & 0. 6049 \\ \cline{2-5} 
                           & 0.001                           & 0. 6057      & 0. 6061     & 0. 6064 \\ \cline{2-5} 
                           & 0.0005                          & 0. 6075     & 0. 6077       & 0. 6084 \\ \cline{2-5} 
                           & 0.0001                          & 0. 6088        & 0. 6092       & 0. 6096 \\ \hline
\end{tabular}
\end{center}
\end{table}

%%%%%%%%%%%%%%%%%%%%%%%%%%%%%%%%%%%%%%%%%%%%%%%%%%%%%%%%%%%%%%

%%%%%%%%%%%%%%%%%%%%%%%%%%%%%%%%%%%%%%%%%%%%%%%%%%%%%%%%
%\textcolor{red}{Table number 5}
\begin{table}[h!]
\begin{center}
	\caption{
	{\bf Hyperparameter tuning for 150 neurons single layer GRU with Spearmans RSFC.}}
	\label{Table:Hyperparameter tuning for 150 neurons single layer GRU with Spearmans RSFC}
		%\resizebox{13cm}{!}{
\begin{tabular}{|c|c|c|c|c|c|}
\hline
\multirow{2}{*}{Optimizer} & \multirow{2}{*}{Learning rate} & \multicolumn{3}{c|}{Batch size} \\ \cline{3-5} 
                           &                                & 4         & 8        & 16    \\ \hline
\multirow{4}{*}{Adam}      & 0.01                            & 0. 6314     & 0. 6304    & 0. 6295 \\ \cline{2-5} 
                           & 0.001                           & 0. 6282       & 0. 6267 &  0. 6246 \\ \cline{2-5} 
                           & 0.0005                          & 0. 6234      & 0. 6228       & 0. 6235  \\ \cline{2-5} 
                           & 0.0001                          & 0. 6238     & \bf{0. 6239}      & 0. 6249 \\ \hline
\multirow{4}{*}{Nadam}     & 0.01                            & 0. 6252        & 0. 6253     & 0. 6251 \\ \cline{2-5} 
                           & 0.001                           & 0. 6254      & 0. 6262       & 0. 6268    \\ \cline{2-5} 
                           & 0.0005                           &0. 6256         & 0. 6250       & 0. 6230 \\ \cline{2-5} 
                           & 0.0001                          & 0. 6192      & 0. 6150       & 0. 6120 \\ \hline
\multirow{4}{*}{Adagrad}   & 0.01                            &0. 6123       & 0. 6130     & 0. 6136 \\ \cline{2-5} 
                           & 0.001                           & 0. 6139      & 0. 6139     & 0. 6141 \\ \cline{2-5} 
                           & 0.0005                          & 0. 6144     & 0. 6146       & 0. 6150 \\ \cline{2-5} 
                           & 0.0001                          & 0. 6154        & 0. 6154       & 0. 6159 \\ \hline
\end{tabular}
\end{center}
\end{table}

%%%%%%%%%%%%%%%%%%%%%%%%%%%%%%%%%%%%%%%%%%%%%%%%%%%%%%%%%%%%%%
%%%%%%%%%%%%%%%%%%%%%%%%%%%%%%%%%%%%%%%%%%%%%%%%%%%%%%%%
%\textcolor{red}{Table number 6}
\begin{table}[h!]
\begin{center}
	\caption{
	{\bf Hyperparameter tuning for 200 neurons single layer GRU with Spearmans RSFC.}}
	\label{Table:Hyperparameter tuning for 200 neurons single layer GRU with Spearmans RSFC}
		%\resizebox{13cm}{!}{
\begin{tabular}{|c|c|c|c|c|c|}
\hline
\multirow{2}{*}{Optimizer} & \multirow{2}{*}{Learning rate} & \multicolumn{3}{c|}{Batch size} \\ \cline{3-5} 
                           &                                & 4         & 8        & 16    \\ \hline
\multirow{4}{*}{Adam}      & 0.01                            & 0. 6379     & 0. 6421    & 0. 6336 \\ \cline{2-5} 
                           & 0.001                           & 0. 6284       & 0. 6273 &  0. 6263 \\ \cline{2-5} 
                           & 0.0005                          & 0. 6251      & 0. 6238       & 0. 6229  \\ \cline{2-5} 
                           & 0.0001                          & 0. 6224     & \bf{0. 6229}      & 0. 6223 \\ \hline
\multirow{4}{*}{Nadam}     & 0.01                            & 0. 6224        & 0. 6230     & 0. 6223 \\ \cline{2-5} 
                           & 0.001                           & 0. 6231      & 0. 6238       & 0. 6237    \\ \cline{2-5} 
                           & 0.0005                           &0. 6241         & 0. 6232       & 0. 6215 \\ \cline{2-5} 
                           & 0.0001                          & 0. 6176      & 0. 6143       & 0. 6112 \\ \hline
\multirow{4}{*}{Adagrad}   & 0.01                            &0. 6121       & 0. 6127     & 0. 6131 \\ \cline{2-5} 
                           & 0.001                           & 0. 6134      & 0. 6136     & 0. 6140 \\ \cline{2-5} 
                           & 0.0005                          & 0. 6141     & 0. 6143       & 0. 6146 \\ \cline{2-5} 
                           & 0.0001                          & 0. 6149        & 0. 6150       & 0. 6154 \\ \hline
\end{tabular}
\end{center}
\end{table}
%%%%%%%%%%%%%%%%%%%%%%%%%%%%%%%%%%%%%%%%%%%%%%%%%%%%%%%%%%%%%%
%%%%%%%%%%%%%%%%%%%%%%%%%%%%%%%%%%%%%%%%%%%%%%%%%%%%%%%%
%\textcolor{red}{Table number 7}
\begin{table}[h!]
\begin{center}
	\caption{
	{\bf Hyperparameter tuning for 250 neurons single layer GRU with Spearmans RSFC.}}
	\label{Table:Hyperparameter tuning for 250 neurons single layer GRU with Spearmans RSFC}
		%\resizebox{13cm}{!}{
\begin{tabular}{|c|c|c|c|c|c|}
\hline
\multirow{2}{*}{Optimizer} & \multirow{2}{*}{Learning rate} & \multicolumn{3}{c|}{Batch size} \\ \cline{3-5} 
                           &                                & 4         & 8        & 16    \\ \hline
\multirow{4}{*}{Adam}      & 0.01                            & 0. 6250     & 0. 6289    & 0. 6252 \\ \cline{2-5} 
                           & 0.001                           & 0. 6213       & 0. 6191 &  0. 6183 \\ \cline{2-5} 
                           & 0.0005                          & 0. 6191      & 0. 6181       & 0. 6183  \\ \cline{2-5} 
                           & 0.0001                          & 0. 6186     & \bf{0. 6179}      & 0. 6190 \\ \hline
\multirow{4}{*}{Nadam}     & 0.01                            & 0. 6193        & 0. 6193     & 0. 6202 \\ \cline{2-5} 
                           & 0.001                           & 0. 6209      & 0. 6210       & 0. 6206    \\ \cline{2-5} 
                           & 0.0005                           &0. 6215         & 0. 6219       & 0. 6206 \\ \cline{2-5} 
                           & 0.0001                          & 0. 6171     & 0. 6144      & 0. 6114 \\ \hline
\multirow{4}{*}{Adagrad}   & 0.01                            &0. 6122       & 0. 6122     & 0. 6131 \\ \cline{2-5} 
                           & 0.001                           & 0. 6130      & 0. 6133     & 0. 6134 \\ \cline{2-5} 
                           & 0.0005                          & 0. 6133     & 0. 6133       & 0. 6136 \\ \cline{2-5} 
                           & 0.0001                          & 0. 6138        & 0. 6140       & 0. 6143 \\ \hline
\end{tabular}
\end{center}
\end{table}

%%%%%%%%%%%%%%%%%%%%%%%%%%%%%%%%%%%%%%%%%%%%%%%%%%%%%%%%%%%%%%
%%%%%%%%%%%%%%%%%%%%%%%%%%%%%%%%%%%%%%%%%%%%%%%%%%%%%%%%%%%%%%
\clearpage
\subsection*{Hyperparameter tuning for single layer GRU on the Partial Correlation's RSFC}
%%%%%%%%%%%%%%%%%%%%%%%%%%%%%%%%%%%%%%%%%%%%%%%%%%%%%%%%%%%%%%

%%%%%%%%%%%%%%%%%%%%%%%%%%%%%%%%%%%%%%%%%%%%%%%%%%%%%%%%
%\textcolor{red}{Table number 1}
\begin{table}[h!]
\begin{center}
	\caption{
	{\bf Hyperparameter tuning for 10 neurons single layer GRU with Partial RSFC.}}
	\label{Table:Hyperparameter tuning for 10 neurons single layer GRU with Partial RSFC}
		%\resizebox{13cm}{!}{
\begin{tabular}{|c|c|c|c|c|c|}
\hline
\multirow{2}{*}{Optimizer} & \multirow{2}{*}{Learning rate} & \multicolumn{3}{c|}{Batch size} \\ \cline{3-5} 
                           &                                & 4         & 8        & 16    \\ \hline
\multirow{3}{*}{Adam}      & 0.01                            & 0. 5271     & 0. 5282    & 0. 5283 \\ \cline{2-5} 
                           & 0.001                           & 0. 5345       & 0. 5409 &  0. 5424 \\ \cline{2-5} 
             
                           & 0.0001                          & 0. 5453       & \bf{0.5468}      & 0. 5459 \\ \hline
\multirow{3}{*}{Nadam}     & 0.01                            & 0. 5459        & 0. 5463       & 0. 5463 \\ \cline{2-5} 
                           & 0.001                           & 0. 5449      & 0. 5438       & 0. 5436    \\ \cline{2-5} 
                           
                           & 0.0001                          & 0. 5415      & 0. 5393       & 0. 5373 \\ \hline
\multirow{3}{*}{Adagrad}   & 0.01                            &0. 5371        & 0. 5368      & 0. 5365 \\ \cline{2-5} 
                           & 0.001                           & 0. 5373      & 0. 5385     & 0. 5393 \\ \cline{2-5} 
                           
                           & 0.0001                          & 0. 5389        & 0. 5394       & 0. 5395 \\ \hline
\end{tabular}
\end{center}
\end{table}

%%%%%%%%%%%%%%%%%%%%%%%%%%%%%%%%%%%%%%%%%%%%%%%%%%%%%%%%%%%%%%
%%%%%%%%%%%%%%%%%%%%%%%%%%%%%%%%%%%%%%%%%%%%%%%%%%%%%%%%
%\textcolor{red}{Table number 2}
\begin{table}[h!]
\begin{center}
	\caption{
	{\bf Hyperparameter tuning for 30 neurons single layer GRU with Partial RSFC.}}
	\label{Table:Hyperparameter tuning for 30 neurons single layer GRU with Partial RSFC}
		%\resizebox{13cm}{!}{
\begin{tabular}{|c|c|c|c|c|c|}
\hline
\multirow{2}{*}{Optimizer} & \multirow{2}{*}{Learning rate} & \multicolumn{3}{c|}{Batch size} \\ \cline{3-5} 
                           &                                & 4         & 8        & 16    \\ \hline
\multirow{3}{*}{Adam}      & 0.01                            & 0. 5171     & 0. 5186    & 0. 5186 \\ \cline{2-5} 
                           & 0.001                           & 0. 5257       & 0. 5317 &  0. 5356 \\ \cline{2-5} 
             
                           & 0.0001                          & 0. 5382       & \bf{0. 5385}      & 0. 5420 \\ \hline
\multirow{3}{*}{Nadam}     & 0.01                            & 0. 5446        & 0. 5458       & 0. 5463 \\ \cline{2-5} 
                           & 0.001                           & 0. 5459      & 0. 5453       & 0. 5441    \\ \cline{2-5} 
                           
                           & 0.0001                          & 0. 5428      & 0. 5424       & 0. 5394 \\ \hline
\multirow{3}{*}{Adagrad}   & 0.01                            &0. 5385        & 0. 5377      & 0. 5366 \\ \cline{2-5} 
                           & 0.001                           & 0. 5371      & 0. 5376     & 0. 5380 \\ \cline{2-5} 
                           
                           & 0.0001                          & 0. 5389        & 0. 5392       & 0. 5401 \\ \hline
\end{tabular}
\end{center}
\end{table}

%%%%%%%%%%%%%%%%%%%%%%%%%%%%%%%%%%%%%%%%%%%%%%%%%%%%%%%%%%%%%%
%%%%%%%%%%%%%%%%%%%%%%%%%%%%%%%%%%%%%%%%%%%%%%%%%%%%%%%%
%\textcolor{red}{Table number 3}
\begin{table}[h!]
\begin{center}
	\caption{
	{\bf Hyperparameter tuning for 50 neurons single layer GRU with Partial RSFC.}}
	\label{Table:Hyperparameter tuning for 50 neurons single layer GRU with Partial RSFC}
		%\resizebox{13cm}{!}{
\begin{tabular}{|c|c|c|c|c|c|}
\hline
\multirow{2}{*}{Optimizer} & \multirow{2}{*}{Learning rate} & \multicolumn{3}{c|}{Batch size} \\ \cline{3-5} 
                           &                                & 4         & 8        & 16    \\ \hline
\multirow{3}{*}{Adam}      & 0.01                            & 0. 5321     & 0. 5300    & 0. 5267 \\ \cline{2-5} 
                           & 0.001                           & 0. 5323       & 0. 5369 &  0. 54012 \\ \cline{2-5} 
             
                           & 0.0001                          & 0. 5420       & \bf{0. 5432}      & 0. 5429 \\ \hline
\multirow{3}{*}{Nadam}     & 0.01                            & 0. 5446        & 0. 5465       & 0. 5483 \\ \cline{2-5} 
                           & 0.001                           & 0. 5476      & 0. 5470       & 0. 5468    \\ \cline{2-5} 
                           
                           & 0.0001                          & 0. 5439      & 0. 5418       & 0. 5388 \\ \hline
\multirow{3}{*}{Adagrad}   & 0.01                            &0. 5386        & 0. 5380      & 0. 5371 \\ \cline{2-5} 
                           & 0.001                           & 0. 5377      & 0. 5384     & 0. 5393 \\ \cline{2-5} 
                           
                           & 0.0001                          & 0. 5403        & 0. 5412       & 0. 5413 \\ \hline
\end{tabular}
\end{center}
\end{table}

%%%%%%%%%%%%%%%%%%%%%%%%%%%%%%%%%%%%%%%%%%%%%%%%%%%%%%%%%%%%%%

%%%%%%%%%%%%%%%%%%%%%%%%%%%%%%%%%%%%%%%%%%%%%%%%%%%%%%%%
%\textcolor{red}{Table number 4}
\begin{table}[h!]
\begin{center}
	\caption{
	{\bf Hyperparameter tuning for 100 neurons single layer GRU with Partial RSFC.}}
	\label{Table:Hyperparameter tuning for 100 neurons single layer GRU with Partial RSFC}
		%\resizebox{13cm}{!}{
\begin{tabular}{|c|c|c|c|c|c|}
\hline
\multirow{2}{*}{Optimizer} & \multirow{2}{*}{Learning rate} & \multicolumn{3}{c|}{Batch size} \\ \cline{3-5} 
                           &                                & 4         & 8        & 16    \\ \hline
\multirow{3}{*}{Adam}      & 0.01                            & 0. 5343     & 0. 5314    & 0. 5321 \\ \cline{2-5} 
                           & 0.001                           & 0. 5371       & 0. 538 &  0. 5392 \\ \cline{2-5} 
             
                           & 0.0001                          & 0. 5431       & \bf{0. 5443}      & 0. 5435 \\ \hline
\multirow{3}{*}{Nadam}     & 0.01                            & 0. 5447        & 0. 5460       & 0. 5473 \\ \cline{2-5} 
                           & 0.001                           & 0. 5471      & 0. 5466       & 0. 5457    \\ \cline{2-5} 
                           
                           & 0.0001                          & 0. 5425      & 0. 5416       & 0. 5396 \\ \hline
\multirow{3}{*}{Adagrad}   & 0.01                            &0. 5389        & 0. 5387      & 0. 5374 \\ \cline{2-5} 
                           & 0.001                           & 0. 5380      & 0. 5383     & 0. 5388 \\ \cline{2-5} 
                           
                           & 0.0001                          & 0. 5399        & 0. 5408       & 0. 5408 \\ \hline
\end{tabular}
\end{center}
\end{table}

%%%%%%%%%%%%%%%%%%%%%%%%%%%%%%%%%%%%%%%%%%%%%%%%%%%%%%%%%%%%%%

%%%%%%%%%%%%%%%%%%%%%%%%%%%%%%%%%%%%%%%%%%%%%%%%%%%%%%%%
%\textcolor{red}{Table number 5}
\begin{table}[h!]
\begin{center}
	\caption{
	{\bf Hyperparameter tuning for 150 neurons single layer GRU with Partial RSFC.}}
	\label{Table:Hyperparameter tuning for 150 neurons single layer GRU with Partial RSFC}
		%\resizebox{13cm}{!}{
\begin{tabular}{|c|c|c|c|c|c|}
\hline
\multirow{2}{*}{Optimizer} & \multirow{2}{*}{Learning rate} & \multicolumn{3}{c|}{Batch size} \\ \cline{3-5} 
                           &                                & 4         & 8        & 16    \\ \hline
\multirow{3}{*}{Adam}      & 0.01                            & 0. 5393     & 0. 53    & 0. 5279 \\ \cline{2-5} 
                           & 0.001                           & 0. 5320       & 0. 5360 &  0. 5389 \\ \cline{2-5} 
             
                           & 0.0001                          & 0. 5390       & \bf{0. 5409}      & 0. 5422 \\ \hline
\multirow{3}{*}{Nadam}     & 0.01                            & 0. 5432        & 0. 5442       & 0. 5454 \\ \cline{2-5} 
                           & 0.001                           & 0. 5451      & 0. 5456       & 0. 5447    \\ \cline{2-5} 
                           
                           & 0.0001                          & 0. 5424      & 0. 5411       & 0. 5385 \\ \hline
\multirow{3}{*}{Adagrad}   & 0.01                            &0. 5382        & 0. 5375      & 0. 5370 \\ \cline{2-5} 
                           & 0.001                           & 0. 5377      & 0. 5382     & 0. 5388 \\ \cline{2-5} 
                           
                           & 0.0001                          & 0. 5393        & 0. 5396       & 0. 5397 \\ \hline
\end{tabular}
\end{center}
\end{table}

%%%%%%%%%%%%%%%%%%%%%%%%%%%%%%%%%%%%%%%%%%%%%%%%%%%%%%%%%%%%%%

%%%%%%%%%%%%%%%%%%%%%%%%%%%%%%%%%%%%%%%%%%%%%%%%%%%%%%%%
%\textcolor{red}{Table number 6}
\begin{table}[h!]
\begin{center}
	\caption{
	{\bf Hyperparameter tuning for 200 neurons single layer GRU with Partial RSFC.}}
	\label{Table:Hyperparameter tuning for 200 neurons single layer GRU with Partial RSFC}
		%\resizebox{13cm}{!}{
\begin{tabular}{|c|c|c|c|c|c|}
\hline
\multirow{2}{*}{Optimizer} & \multirow{2}{*}{Learning rate} & \multicolumn{3}{c|}{Batch size} \\ \cline{3-5} 
                           &                                & 4         & 8        & 16    \\ \hline
\multirow{3}{*}{Adam}      & 0.01                            & 0. 5464     & 0. 5286    & 0. 5229 \\ \cline{2-5} 
                           & 0.001                           & 0. 5296       & 0. 5337 &  0. 5369 \\ \cline{2-5} 
             
                           & 0.0001                          & 0. 5384       & \bf{0. 5396}      & 0. 5402 \\ \hline
\multirow{3}{*}{Nadam}     & 0.01                            & 0. 5432        & 0. 5453       & 0. 5461 \\ \cline{2-5} 
                           & 0.001                           & 0. 5466      & 0. 5461       & 0. 5460    \\ \cline{2-5} 
                           
                           & 0.0001                          & 0. 5444      & 0. 5411       & 0. 5378 \\ \hline
\multirow{3}{*}{Adagrad}   & 0.01                            &0. 5378        & 0. 5373      & 0. 5366 \\ \cline{2-5} 
                           & 0.001                           & 0. 5369      & 0. 5374     & 0. 5375 \\ \cline{2-5} 
                           
                           & 0.0001                          & 0. 5380        & 0. 5384       & 0. 5388 \\ \hline
\end{tabular}
\end{center}
\end{table}

%%%%%%%%%%%%%%%%%%%%%%%%%%%%%%%%%%%%%%%%%%%%%%%%%%%%%%%%%%%%%%

%%%%%%%%%%%%%%%%%%%%%%%%%%%%%%%%%%%%%%%%%%%%%%%%%%%%%%%%
%\textcolor{red}{Table number 7}
\begin{table}[h!]
\begin{center}
	\caption{
	{\bf Hyperparameter tuning for 250 neurons single layer GRU with Partial RSFC.}}
	\label{Table:Hyperparameter tuning for 250 neurons single layer GRU with Partial RSFC}
		%\resizebox{13cm}{!}{
\begin{tabular}{|c|c|c|c|c|c|}
\hline
\multirow{2}{*}{Optimizer} & \multirow{2}{*}{Learning rate} & \multicolumn{3}{c|}{Batch size} \\ \cline{3-5} 
                           &                                & 4         & 8        & 16    \\ \hline
\multirow{3}{*}{Adam}      & 0.01                            & 0. 545     & 0. 5357    & 0. 535 \\ \cline{2-5} 
                           & 0.001                           & 0. 5379       & 0. 5410 &  0. 5408 \\ \cline{2-5} 
             
                           & 0.0001                          & 0. 5431       & \bf{0. 5441}      & 0. 5453 \\ \hline
\multirow{3}{*}{Nadam}     & 0.01                            & 0. 5474        & 0. 5491       & 0. 5511 \\ \cline{2-5} 
                           & 0.001                           & 0. 5514      & 0. 5499       & 0. 5493    \\ \cline{2-5} 
                           
                           & 0.0001                          & 0. 5462      & 0. 5437       & 0. 5410 \\ \hline
\multirow{3}{*}{Adagrad}   & 0.01                            &0. 5408        & 0. 5396      & 0. 5385 \\ \cline{2-5} 
                           & 0.001                           & 0. 5391      & 0. 5398     & 0. 5401 \\ \cline{2-5} 
                           
                           & 0.0001                          & 0. 5406        & 0. 5410       & 0. 5413 \\ \hline
\end{tabular}
\end{center}
\end{table}

%%%%%%%%%%%%%%%%%%%%%%%%%%%%%%%%%%%%%%%%%%%
\end{appendices}
%%===========================================================================================%%
%\bibliographystyle{amsplain} % basic style, author-year citations
%\bibliography{Elseiver_ASDProj}
%\label{reference_section}
\bibliography{sn-bibliography}% common bib file
%% if required, the content of .bbl file can be included here once bbl is generated
%%\input sn-article.bbl

%% Default %%
%%\input sn-sample-bib.tex%

\end{document}